\documentclass{article}
\usepackage[utf8]{inputenc}
\usepackage{rotating}
\usepackage{csquotes}
\usepackage{authblk}
\usepackage{xr}

\title{\new{SSRCA: a novel machine learning pipeline to perform sensitivity analysis for agent-based models}}

\date{\today}

\author[1]{Edward H. Rohr}

\author[2]{John T. Nardini}

\affil[1]{Department of Mathematics, Tufts University, Medford, MA, 02155, USA.}

\affil[2]{Department of Mathematics and Statistics, The College of New Jersey, Ewing, NJ, 08628, USA. \\
nardinij@tcnj.edu}

\usepackage{makecell}
\usepackage[square,sort,comma,numbers]{natbib}
\usepackage{multirow}

\usepackage{graphicx}
\usepackage{amsmath}
\usepackage{geometry}
\usepackage{units}
\usepackage{amsfonts}
\usepackage{amssymb}
\usepackage{pifont}
\usepackage[title]{appendix}%
\usepackage{pdflscape}
\usepackage[dvipsnames]{xcolor}
\usepackage{hyperref}
\usepackage{cleveref}
\usepackage{enumitem}
\usepackage{bm}
\usepackage{hyperref}
\usepackage{verbatim}
\usepackage{ulem}
\usepackage[utf8]{inputenc}
\usepackage[T1]{fontenc}

\newcommand{\new}[1]{#1}
\newcommand{\old}[1]{}

\newcommand{\C}{c(\boldsymbol{x},t)}
\newcommand{\p}{\boldsymbol{p}}
\newcommand{\E}{\text{E}}
\newcommand{\Var}{\text{Var}}

\usepackage[ruled,vlined]{algorithm2e}
\usepackage{todonotes}
\usepackage{comment}

\usepackage{xr}
\usepackage{cleveref}

\begin{document}

\maketitle

\begin{abstract}
Agent-based models (ABMs) are widely used in biology to understand how individual actions scale into emergent population behavior. Modelers employ sensitivity analysis (SA) algorithms to quantify input parameters' impact on model outputs, however, it is hard to perform SA for ABMs due to their computational and complex nature. In this work, we develop the \textbf{S}imulate, \textbf{S}ummarize, \textbf{R}educe, \textbf{C}luster, and \textbf{A}nalyze (SSRCA) methodology, a machine-learning based pipeline designed to facilitate SA for ABMs. In particular, SSRCA can achieve the following tasks for ABMS: 1) identify sensitive model parameters, 2) reveal common output model patterns, and 3) determine which input parameter values generate these patterns. We use an example ABM of tumor spheroid growth to showcase how SSRCA identifies four common patterns from the ABM and the parameter regions that generate these outputs. \new{Additionally, we compare the SA results between SSRCA and the popular Sobol' Method and find that SSRCA's identified sensitive parameters are robust to the choice of model descriptors while Sobol's are not.} This analysis could streamline data-driven tasks, such as parameter estimation, for ABMs by reducing parameter space. While we highlight these results with an ABM on tumor spheroid formation, the SSRCA Methodology is broadly applicable to biological ABMs.

\end{abstract}

\section{Introduction}\label{sec:introduction}

Mathematical models provide an efficient and cost-effective approach to generate hypotheses for \new{biological processes} and optimally design experiments \cite{byrne_dissecting_2010, wallace_properties_2013}. Common approaches for simulating \emph{in-silico} \new{experiments} include continuum models \cite{greenspan_models_1972,MCELWAIN1977267,ward_mathematical_1997,10.1093/imammb/18.2.131,Grimes,murphy_growth_2023} and cellular automata \cite{MALLET2006334,bruningk_cellular_2019,messina_hybrid_2023}. Agent-based models (ABMs) are another popular modeling choice due to their ability to capture the discrete and stochastic nature of \new{biological processes} \cite{10.1371/journal.pcbi.1006469, 10.1371/journal.pcbi.1007961,Cooper2020,MALLET2006334,rigid_body}. In an ABM, modelers simulate pre-defined individual-level rules to determine how they translate into collective behavior. For example, Klowss et al. developed a hybrid ABM that consists of both discrete tumor cells and a continuous variable representing a nutrient \new{to imitate tumor spheroid experiments} \cite{4D_Tumor_Model}. This model includes rules dictating cell migration, death, and transition between cell cycle states. Model simulations serve to precisely predict how experimental conditions and/or design choices (including initial spheroid size, kinetic parameter rates, \new{experimental duration}, etc.) impact \new{the final model outcomes}. In another study, Rocha et al. demonstrated that their ABM could be used as a cancer patient digital twin to develop patient-specific models for precision medicine treatment plans \cite{l_rocha_multiscale_2024}.

Despite the wide use of ABMs, their analysis presents many challenges; ABMs typically require many parameter values, are not amenable to extensive simulation due to long simulation times, and their outputs are complex and noisy. These difficulties hinder the application of standard modeling techniques for ABMs, including parameter estimation, uncertainty quantification, and sensitivity analysis. As one example, consider the Klowss Model from \cite{4D_Tumor_Model}: There are 25 model parameters: 13 can be estimated from biological experiments, 2 can be determined from numerical experiments, leaving 10 unknown parameters. Impressively, the authors in this study found values for all 10 parameters and demonstrated excellent agreement between their \emph{in-silico} simulations and \emph{in-vitro} spheroid experiments using human primary melanoma cells. However, there is no guarantee that the same values would generalize to data from other cell lines. Estimating all 10 parameter values from new data would require exhaustively searching the 10-dimensional parameter space, which is infeasible due to long simulation times and the curse of dimensionality \cite{fernandez-martinez_curse_2020}.  In addition to parameter estimation, these obstacles inhibit our ability to perform \emph{feature mapping}, which links the model's input parameter values to its output patterns.

Parameter sensitivity analysis (SA) is an area of research to identify a model's most influential parameters. Many aspects of the modeling process are encompassed in SA, including quantifying the uncertainty of model outputs, determining influential input parameters, and feature mapping \cite{ten_broeke_which_2016,bergman_efficient_2025}. Modelers can fix the insensitive parameters to reduce the dimensionality of the input parameter space while still capturing the main model behavior. There are many different approaches to performing SA \cite{dellino_review_2015,pianosi_sensitivity_2016}. For example, the Morris one-at-a-time (MOAT) method is a computationally-efficient algorithm that varies one parameter at a time. This simple approach ignores interactions between parameters, however, and can thus generate misleading results for complex models. The Sobol' Method is a global SA (GSA) approach that ranks parameters based on their contribution to the model's variance using computed sensitivity index values. These Sobol' index values incorporate parameter interactions, but this added information comes at the cost of higher computational expenses. Regression-based methods, such as partial rank correlation coefficient, are another popular SA method in which one fits a regression model to predict the model's output values from its input parameters. While regression-based  approaches can be used to perform feature mapping, it can be challenging to determine a suitable regression model for nonlinear models \cite{ligmann-zielinska_one_2020}.

\begin{figure}[h!]
    \centering
    \includegraphics[width=0.85\linewidth]{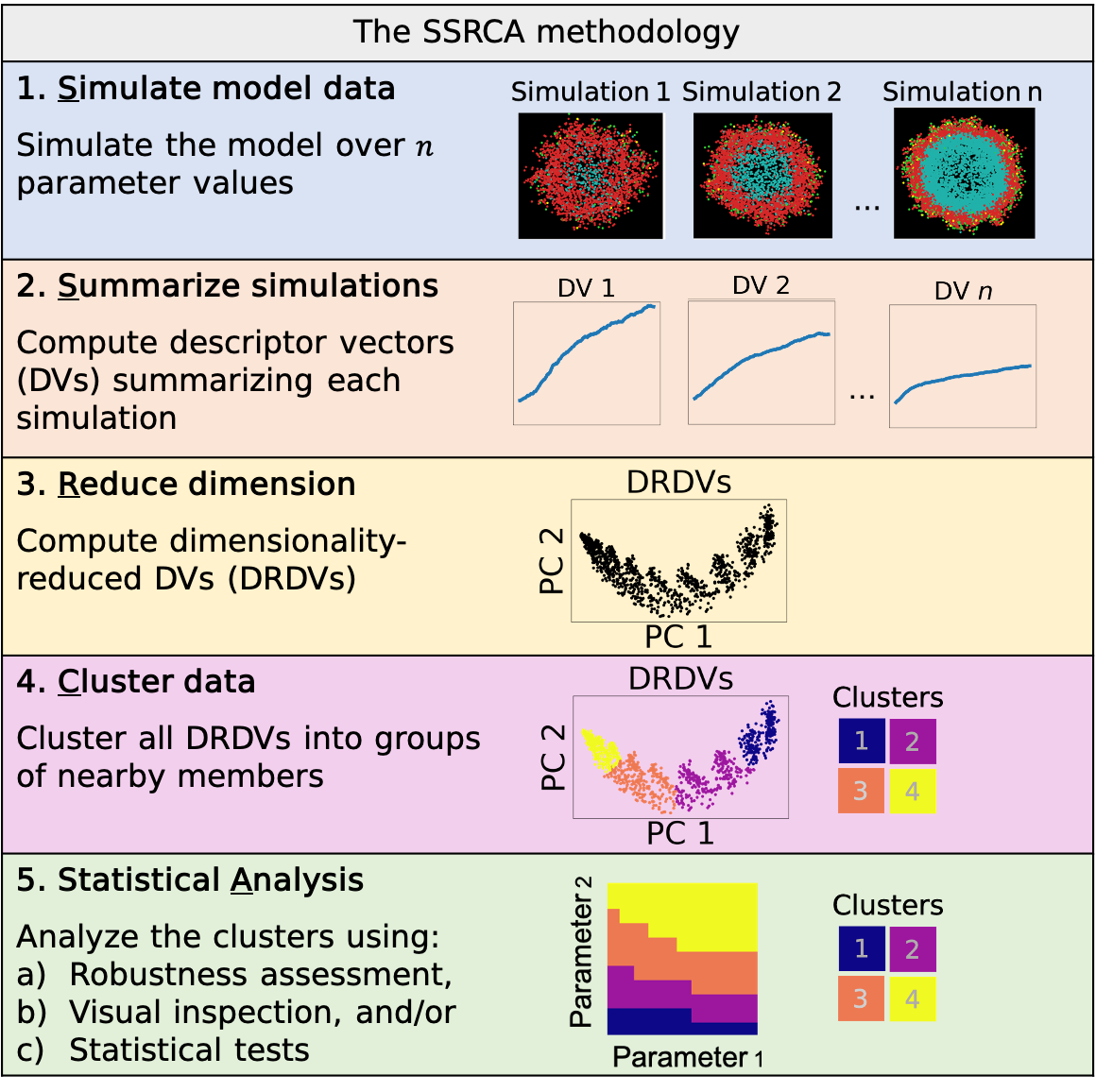}
    \caption{The SSRCA Methodology pipeline. The SSRCA sensitivity analysis pipeline consists of 5 steps. \uline{1. Simulate model data:} Simulate a large number of model simulations from sampled parameter values to generate a dataset. \uline{2. Summarize simulations:} Compute uniform length descriptor vectors (DVs) to summarize each model simulation. \uline{3. Reduce dimension:} Perform dimensionality reduction to reduce each DV to a dimensionality-reduced DV (DRDV). \uline{4. Cluster data:} Cluster all DRDVs into $k$ distinct groups. \uline{5. statistical Analysis:} Analyze the clusterings and their constituent parameter samples by a) assessing cluster robustness, b) visually inspecting each cluster's data, and/or performing statistical tests to infer cluster differences.}
    \label{fig:methods_pipeline}
\end{figure}

Despite the wide and successful use of SA methods, their application to ABMs has been limited. For example, feature mapping is challenging to perform for an ABM because nearby parameter values can yield divergent patterns. Broeke et al. assessed the performance of MOAT and the Sobol' Method for ABMs, and found that the simple MOAT approach can be used for preliminary feature mapping analysis, but that the Sobol' Method cannot because its analyses are computed over the entire parameter space instead of considering distinct parameter regions \cite{ten_broeke_which_2016}. Other challenges for performing SA for ABMs arise due to their time-varying, nonlinear, computational, and multi-scale natures \cite{bergman_modeling_2021,bergman_efficient_2025,ligmann-zielinska_one_2020}. Instead of applying standard GSA methods to ABMs, there is an urgent need for the development of novel ABM-specific SA approaches to fully harness the potential of ABMs for modeling complex phenomena. Towards this end, we introduce the \textbf{S}imulate, \textbf{S}ummarize, \textbf{R}educe, \textbf{C}luster, and \textbf{A}nalyze (SSRCA, pronounced ``circa'') methodology to perform GSA for ABMs \new{(Figure \ref{fig:methods_pipeline})}. An advantage of this approach over other GSA algorithms, such as the Sobol' indices method, is that it can both identify sensitive parameters and perform feature mapping. In particular, SSRCA can achieve the following tasks: 1) identifies sensitive model parameters, 2) reveals common output model patterns, and 3) determines which parameter values generate these patterns. 

\new{We focus our application of the SSRCA Methodology on the Klowss Model from \cite{4D_Tumor_Model}. This 3-dimensional model imitates tumor spheroid experiments, which contain high levels of intratumoral heterogeneity. For example, cells in oxygen rich areas will rapidly divide and increase in number, whereas those in hypoxic regions may become quiescent or die. The \textit{fluorescent ubiquitination-based cell cycle indicator} (FUCCI) imaging technique allows experimentalists to measure and quantify population heterogeneity by tracking individual cells' progression through the cell cycle in real-time \cite{koledova_real-time_2017,FUCCI,Vittadello}. Simulations of the Klowss Model  capture many realistic facets of \textit{in vitro} tumor spheroid growth: an outer proliferating shell, an inner growth-arrested shell, and a necrotic core \cite{murphy_designing_2022}. In this study, we apply the SSRCA Methodology to a 2-dimensional version of the Klowss Model from \cite{4D_Tumor_Model} using two simulated datasets. For a small dataset (generated by varying 2 model parameters), SSRCA directly links parameter values to one of four typical model patterns quantifying the sizes of the proliferating shell and necrotic core. For a larger dataset (generated by varying 10 model parameters), SSRCA identifies four sensitive model parameters, each of which is involved in cell cycle entry or death, and the parameter distributions underlying the common model patterns. These results suggest that cell cycle entry and death are the most impactful processes involved in tumor spheroid formation, and SSRCA reveals the possible model outputs when varying these sensitive parameters. We highlight the SSRCA Methodology's capability for performing GSA on a tumor spheroid ABM in this work, however, this approach is broadly relevant to biological ABMs with applications in intracellular protein dynamics, disease spread, and  ecology \cite{ciocanel_simulated_2022,kim_simulation_2022,bernoff_agent-based_2020}.}

\section{Methods}\label{sec:methods}

In this section, we review the Klowss Model from \cite{4D_Tumor_Model} in Section \ref{sec:model}, detail the SSRCA Methdology in Section \ref{sec:pipeline}, describe how we re-parameterize the Klowss Model in 2 spatial dimensions in Section \ref{sec:reparameterization}, and review the Sobol' Method in Section \ref{sec:sobol_method}.

\subsection{The Klowss Model of tumor spheroid growth}\label{sec:model}

\new{Many mathematical models of tumor spheroid dynamics have been developed previously \cite{karolak_towards_2018}, with most capturing key characteristics of \emph{in vitro} tumor spheroid growth: the formation of a necrotic core in the spheroid's center and an outer proliferating ring due to nutrient diffusion and consumption \cite{murphy_designing_2022}. Within these modeling studies, many different modeling approaches have been implemented to describe tumor cell locations or densities over time, including cellular automata \cite{MALLET2006334, messina_hybrid_2023, 10.1371/journal.pcbi.1006469, bruningk_cellular_2019, song_chen_agent-based_2004}, off-lattice ABMs \cite{10.1371/journal.pcbi.1007961, 4D_Tumor_Model, milotti_emergent_2010}, Voronoi-latticed ABMs \cite{cleri_agent-based_2019, jagiella_inferring_2016}, or differential equation models \cite{greenspan_models_1972, murphy_designing_2022, murphy_growth_2023}. In this study, we choose to focus our application of the SSRCA methodology on the off-lattice Klowss Model from \cite{4D_Tumor_Model} for two key reasons. First, this model has already been calibrated to \emph{in vitro} experimental data previously, which indicates it captures realistic biological behavior. Second, this model is of moderate complexity with 25 parameters (10 of which were assumed in \cite{4D_Tumor_Model} without biological justification). This provides us with an opportunity to determine which of these 10 assumed parameters most significantly impact the model's outputs and to explore how the model's dynamics change when varying the model's sensitive parameters. Even though a 3-dimensional model was presented in \cite{4D_Tumor_Model}, we simulate the model in 2 spatial dimensions to ease the computational requirements; see Section \ref{sec:reparameterization} for more details.}

\begin{figure}
\centering
\includegraphics[width=.95\linewidth]{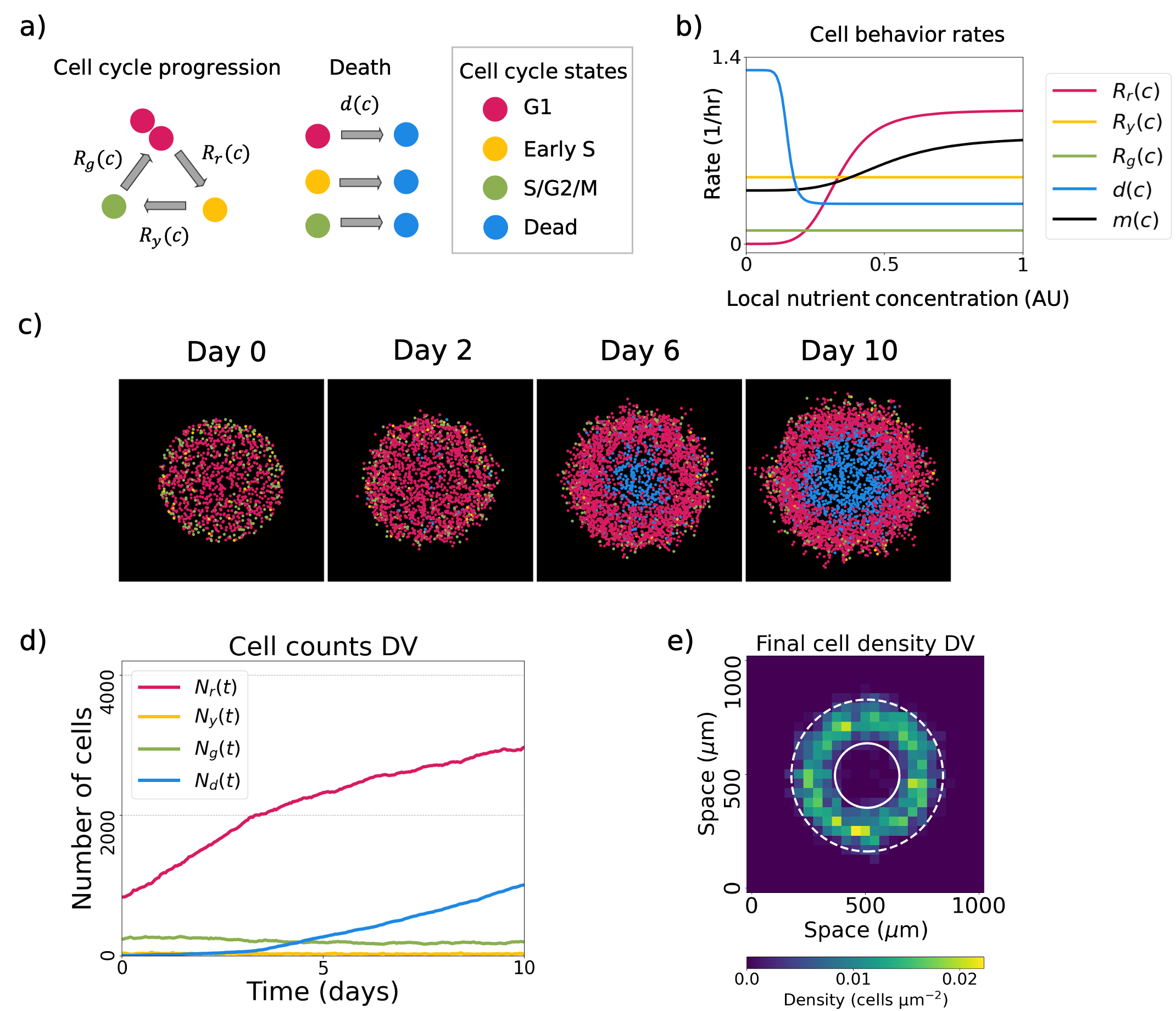}
\caption{Klowss model overview and associated descriptor vectors. (a)  The cell cycle is modeled as follows: agents \new{in the G1 state} transition to \new{the early S state} with rate $R_r(c)$, \new{agents in the early S state} transition to \new{the S/G2/M state} with rate $R_y(c)$, agents \new{in the S/G2/M state} undergo mitosis with rate $R_g(x)$ in which they become red and create a daughter red cells \new{in the G1 state}; the daughter cell is placed one cell diameter away in a randomly chosen direction. All agents die with rate $d(c)$ and change their cell cycle status to dead. Not shown: All agents migrate with rate $m(c)$, and move the length of a cell diameter in a randomly chosen direction. (b) All model rates depend on the nutrient available \new{(in arbitrary units (AU))} at the agent's physical location. \new{The depicted rates qualitatively illustrate cells' behavior in the model but are not the precise rates used for the simulations. } (c)  Snapshots at times t = 0, 2, 6, and 10 days from the baseline simulation. d) The Cell counts DV for the baseline simulation. e) The Final cell density DV for the baseline simulation. The solid (dashed) circle has a radius of 137 $\mu m$ (322 $\mu m$). }
\label{fig:model_overview}
\end{figure}

We provide details on the Klowss Model and our implementation in Appendix \ref{app:model_overview} \new{and the baseline parameter values used in Table \ref{table: parameters perturbed}}. \new{Briefly, the model contains $N(t)$ agents. Each agent may be in one of three living states: G1, early S, or G2/S/M (Figure \ref{fig:model_overview}(a)).  All living agents perform rules on migration, death, and cell cycle progression; agents undergo mitosis by creating a daughter cell \new{in the G1 state} when transitioning from \new{the S/G2/M state to the G1 state}
 (Figure \ref{fig:model_overview}(a)). Additionally, cells can transition from any of the three living states into the dead state (Figure \ref{fig:model_overview}(a)).} The kinetic rates for each rule depend on a continuous variable, $\C$, such as oxygen (Figure \ref{fig:model_overview}(b)). \new{The rates of cell migration, $m(c)$, and entry into the cell cycle, $R_r(c),$ are both modeled using Hill functions that increase with $\C$; the death rate, $d(c)$, is modeled using a Hill function that decreases with $\C$; and the rates from early S to S/G2M, $R_y(c)$, and for mitosis, $R_g(c)$, are both constant.} The spatiotemporal concentration of $\C$ varies due to diffusion and agent consumption, see Appendix \ref{Nutrient dynamics}. \new{ We describe the initialization process for the model in Appendix \ref{initialization}}

\begin{table}
\centering
\begin{tabular}{|l|c|c|c|c|c|}
\hline
\textbf{Parameter name} & \textbf{Symbol} & \textbf{Base value}&  \textbf{Varying} & \textbf{Minimum} & \textbf{Maximum}\\
Maximum death rate & $d_{max}$ & 2 $\mathrm{h}^{-1}$& yes&1 $\mathrm{h}^{-1}$&4 $\mathrm{h}^{-1}$\\
Minimum death rate & $d_{min}$ & 0.0005 $\mathrm{h}^{-1}$& yes& 0.00025 $\mathrm{h}^{-1}$& 0.001 $\mathrm{h}^{-1}$\\
Maximum migration rate & $m_{max}$ & 0.12 $\mathrm{h}^{-1}$& yes& 0.06 $\mathrm{h}^{-1}$& 0.24 $\mathrm{h}^{-1}$\\
Minimum migration rate & $m_{min}$ & 0.06 $\mathrm{h}^{-1}$& yes& 0.03 $\mathrm{h}^{-1}$& 0.12 $\mathrm{h}^{-1}$\\
Hill function index for arrest & $\eta_1$ & 5& yes& 2.5& 10\\
Hill function index for migration & $\eta_2$ & 5& yes& 2.5& 10\\
Hill function index for death & $\eta_3$ & 15& yes& 7.5& 30\\
Critical arrest concentration & $c_a$ & 0.4& yes& 0.2& 0.8\\
Critical migration concentration & $c_m$ & 0.5& yes& 0.25& 1\\
Critical death concentration & $c_d$ & 0.1 & yes& 0.05& 0.2\\
Initial number of agents & $N(0)$ & 1100 & no & N/A & N/A\\
\new{Cell diameter} & \new{$\mu$} & \new{12 $\mu$m} & \new{no} & \new{N/A} & \new{N/A} \\
Domain length & $L$ & 1000 $\mu\mathrm{m}$ & no & N/A & N/A\\
Initial spheroid radius & $r_o(0)$ & 245 $\mu\mathrm{m}$ & no & N/A & N/A\\
Dispersal/migration distance & $\mu$ & 12 $\mu\mathrm{m}$ & no & N/A & N/A\\
Simulation time & $T$ & 240 $\mathrm{h}$& no & N/A & N/A\\
Maximum G1-eS transition rate & $R_r$ & 0.047 $\mathrm{h}^{-1}$& no & N/A & N/A\\
Constant eS-S/G2/M transition rate & $R_y$ & 0.50 $\mathrm{h}^{-1}$& no & N/A & N/A\\
Constant S/G2/M-G1 transition rate (mitosis) & $R_g$ & 0.062$\mathrm{h}^{-1}$& no & N/A & N/A\\
Number of nodes & $I^2$ & $200^2$ & no & N/A & N/A\\
Steady-state solution interval & $t^*$ & 1 $\mathrm{h}$& no & N/A & N/A\\
Consumption-diffusion ratio & $\alpha$ & 0.01 $\mu\mathrm{m}$ cell$^{-1}$ & no & N/A & N/A\\
\hline
\end{tabular}
\caption{ABM parameters. For all the perturbed parameters, the minimum parameter value is always half the baseline value, while the maximum parameter value is always double the baseline value. \new{All baseline values are from \cite{4D_Tumor_Model}, except $N(0)$ and $\alpha$, which are determined in Section \ref{sec:reparameterization}. For more detailed parameter information, see Section \ref{app:model_overview}.} \label{table: parameters perturbed}}
\end{table}


\subsection{The SSRCA Methodology}\label{sec:pipeline}

We introduce the SSRCA Methodology, which is designed to perform GSA for ABMs (Figure \ref{fig:methods_pipeline}). The SSRCA Methodology pipeline includes 5 steps:

\begin{enumerate}
    \item \textbf{S}imulate model data (Section \ref{sec:model_simulation}),
    \item \textbf{S}ummarize simulations (Section \ref{sec:summarize_simulations}), 
    \item \textbf{R}educe dimension (Section \ref{sec:dimensionality_reduction}), 
    \item \textbf{C}luster data (Section \ref{sec:data_clustering}), and
    \item statistical \textbf{A}nalysis (Section \ref{sec:statistical_analysis}).
\end{enumerate}
\noindent We now detail all five steps of this methodology. All code and simulated data for this study is implemented in Python (version 3.10.11) and publicly available at \href{https://github.com/e-rohr/FUCCI_ABM2D}{https://github.com/e-rohr/FUCCI\_ABM2D}.

\subsubsection{Simulate model data} \label{sec:model_simulation}

We simulate the model over a sample of parameter values from a specified distribution. In this study, we apply the SSRCA Methodology to two simulated datasets to illustrate its performance on \new{both a small and large dataset}. A brief description of each parameter, including which parameters are varied and the minimum and maximum values of all varied parameters, is provided in Table \ref{table: parameters perturbed}. For simplicity, all parameters were varied between half of their baseline value and twice their baseline value.

\noindent \uline{\textbf{\new{The simple 2-parameter dataset}:}} In this dataset, we vary the Hill function index for arrest, $\eta_1$, and the critical arrest concentration, $c_a$, and set all other parameters to their base values from Table \ref{table: parameters perturbed}. 
We vary both $\eta_1$ and $c_a$ over 11 logarithmically spaced values that range over their minimum and maximum values, totaling 121 parameter combinations. We perform 10 model simulations for each combination, resulting in $11\times11\times10=1,210$ total model simulations.

\noindent \uline{\textbf{\new{The large 10-parameter dataset}:}} In this dataset, we vary parameters two at a time while fixing all others at their base values. We consider all ${10 \choose 2} = 45$ pairs of the varying parameters from Table \ref{table: parameters perturbed}.  For each pair, we vary both parameters over 11 logarithmically spaced values that range over their minimum and maximum values, totaling 121 parameter combinations for each pair of parameters. We perform 10 model simulations for each parameter combination, leading to $45\times11\times11\times10=54,450$ total simulations. 

\new{We used The College of New Jersey's Electronic Laboratory for Science and Analysis (ELSA) high performance computer. We computed the large 10-parameter dataset using 450 1-core SLURM jobs, each with 6GB of RAM. Each job ran on a node consisting of a dual Intel(R) Xeon(R) Gold 6130 CPU processer with 192 total GB of RAM.  All 450 jobs completed within 24 hours.}

\subsubsection{Summarize simulations} \label{sec:summarize_simulations}

Recording the locations and cell cycle states of all $N(t)$ agents over time results in prohibitively large datasets because each simulation contains thousands of agents. Furthermore, standard machine learning algorithms require fixed length vectors, but each simulation has a different number of cell counts. For these reasons, we compute fixed length \emph{descriptor vectors} (DVs) to concisely summarize each model simulation. We consider two separate DVs in this study: cell subpopulation counts over time, and the final living cell density over space. Other DVs are possible; the appropriate DV choice depends on the model approach and application \cite{angio_tda,bhaskar_analyzing_2019}.

\noindent \uline{\textbf{Cell counts:}} We count the number of red, yellow, green, and dead cells over time as $N_r(\boldsymbol{t})$, $N_y(\boldsymbol{t})$, $N_g(\boldsymbol{t})$, and $N_d(\boldsymbol{t})$, respectively, for \new{the 241 timepoints at $\boldsymbol{t} = \{0, 1/24, 2/24, \dots, 10 \text{ days}\}$}. For the final DV, we concatenate these four vectors together into a vector of length $4\times241=964$.

\noindent \uline{\textbf{Final Cell Density:}} We create a two-dimensional histogram of living agent (those colored red, yellow, and green) locations at the final timepoint \new{($t=10$ days)} using binning values $\boldsymbol{x}^\text{bin}=\boldsymbol{y}^\text{bin}=\{0, 40, 80, \dots, 1000 \mu \text{m}\}$. We divide each histogram value by the binned area to estimate the spatial cell density. We vectorize each 2-dimensional density of size $25\times25$ into a final vector of length 625.

\subsubsection{Reduce dimension} \label{sec:dimensionality_reduction}

We use principal components analysis (PCA) to reduce the dimension of each DV, resulting in \emph{dimensionality-reduced DVs} (DRDVs). We use scree plots to display the explained variance of each principal component and determine a suitable dimension for the DRDVs \cite{james_introduction_2023}. We use the python package \textbf{Scikit-learn} (version 1.4.2) to perform PCA \new{and data scaling}.

\new{Prior to performing PCA, we perform data scaling to standardize the DVs as follows. We do not perform data scaling for the Final cell density DV because different levels of variance are expected between different pixels. For example, some pixel values are zero for all simulations and have a variance of zero, while others will vary significantly between simulations and have large variance values. For the Cell counts DV, we first standardize each cell counts vector (e.g., $N_r(t), N_y(t), N_g(t), \text{ and } N_d(t)$) to have a variance of one (over all timepoints and data samples) before concatenating them together into the final DV. This is done to ensure that the PCA results are not corrupted by vectors with large magnitudes.} 

\subsubsection{Data clustering} \label{sec:data_clustering}

We use $k$-means clustering \cite{kmeans}, an unsupervised machine learning method, to group each DRDV into one of $k$ clusters. The $k$-means algorithm clusters $d$-dimensional data by identifying $k$ mean vectors, $\bar{x}_1,\bar{x}_2,\dots,\bar{x}_k \in \mathbb{R}^d$. Each data sample is assigned into the cluster whose mean it is closest to, using Euclidean distance. More information on how the mean vectors are found is provided in \cite{kmeans}. We use the elbow method to determine the number of clusters to use for each dataset \cite{james_introduction_2023}. We use the python package \textbf{Scikit-learn} (version 1.4.2) to perform clustering. 

\subsubsection{Statistical Analysis} \label{sec:statistical_analysis}

We perform a robustness assessment, visual inspection, and statistical tests to inspect and analyze the clustered data.

\noindent \uline{\textbf{Robustness assessment:}} We test the robustness of the clustering assignments by computing \emph{out-of-sample (OOS) \new{consistency} scores}. We achieve this by labeling each parameter value according to which cluster most of its samples are placed into. A high (low) OOS \new{consistency} score suggests that many simulations from the same parameter value will be placed into the same cluster (different clusters). 

Computing OOS \new{consistency} scores begins by splitting the 10 DRDVs from each parameter sample into 5 training and 5 testing DRDVs. We perform data clustering using the training DRDVs from all parameter values, and then label each parameter value according to the most common cluster assignments from its training DRDVs. To compute the OOS \new{consistency}, we assign a ``ground truth'' label to each testing DRDV based on its parameter value's label. We create a ``predicted'' label for each testing DRDV by running it through the trained clustering process. The OOS \new{consistency} score is then given by the percentage of agreeing ``ground truth'' and ``predicted'' labels. \new{To avoid data leakage, the training and testing datasets are standardized separately, and only the training data is used to determine the principal components for PCA. The testing data is then projected onto the principal components determined from the training data.}

\noindent \uline{\textbf{Visual inspection:}} Each cluster from Section \ref{sec:data_clustering} contains DRDVs that were created over many parameters values. We use visualization to better understand which parameter values generate each cluster. When varying a small number of parameters (such as 2 or 3), we inspect a \emph{parameter partition plot}, which colors each parameter value according to its cluster labeling \cite{angio_tda}. When varying a larger number of parameters, we create \emph{ridgeline plots} for each cluster that depict the distribution of parameter samples within each cluster. Each parameter distribution is estimated using kernel density estimation \cite{thrun_analyzing_2020}. \new{To aid readers in interpreting ridgeline plots, we provide an illustrative example in Appendix \ref{app:ridgeline} and Supplementary Figure \ref{fig:supp_ridgeline_interpretation}. Briefly, sensitive parameters will have restricted domains and distinct distributions between clusters, whereas insensitive parameters will have wide domains and similar distributions between clusters. The heights of all distributions should be ignored in a ridgeline plot because all distributions are normalized to have a uniform height.  } We use the python package \textbf{Joypy} (version 0.2.4) to create ridgeline plots.

\noindent \uline{\textbf{Statistical tests:}} For datasets generated by varying many parameters, we use the discrete two-sample Kolmogorov-Smirnov statistical test to evaluate whether the parameter distributions differ across clusters. \new{We choose the Kolomogorov-Smirnov test in place of other distribution-comparison tests (e.g., Mann-Whitney U test) due to its wide usage in biological studies and because it compares the entire distributions instead of only their mean or median locations. } For each varied parameter, we compare its samples between all ${k \choose 2}$ cluster  pairs. Each comparison produces a $p$-value; low p-values suggest the parameter samples from the two clusters may be drawn from different distributions. We classify a parameter as impactful if the majority of these pairwise comparisons yield $p$-values below 0.05. We modified code from the \textbf{ksdisc} python package to perform statistical tests.

\subsection{Re-parameterizing the 2-dimensional Klowss Model}\label{sec:reparameterization}

\new{The Klowss Model from \cite{4D_Tumor_Model} is 3-dimensional in space ($\boldsymbol{x}=(x,y,z)$), however, computing the model in all 3 spatial dimensions across many parameter values is computationally prohibitive due to long simulation times. We instead simulate the 2-dimensional ($\boldsymbol{x}=(x,y)$) Klowss model, corresponding to the equatorial cross section of the tumor spheroid at $z=0$. This reduction in dimension is justified by the radially symmetric nature of the model \cite{4D_Tumor_Model}. }

\new{To account for this reduction, we modified two baseline parameter values: the number of starting agents, $N(0)$, and the ratio between nutrient consumption ($\kappa$) and diffusion ($D$), given by $\alpha=\kappa/D$. All other baseline values are set to those provided in \cite{4D_Tumor_Model}, which we present in Table \ref{table: parameters perturbed}. We compared the resulting simulations of the 2-dimensional model to the $z=0$ cross section of the full 3-dimensional Klowss Model. We used the Cell counts descriptor vector (see Section \ref{sec:summarize_simulations}) to summarize each simulation. From a small range of values for $N(0)$ and $\alpha$, we found that setting $N(0)=1100$ leads to good agreement on the number of cells in the G1 state for both models for early timepoints (Supplementary Figure \ref{fig:supp_determine_N}), and setting $\alpha=0.01$ leads to good agreement between the two models for most cell states during later timepoints, although the 3-dimensional model contains more dead cells than the 2-dimensional model (Supplementary Figure \ref{fig:supp_determine_alpha}).}

\subsection{The Sobol' GSA Methodology} \label{sec:sobol_method}

The Sobol' Method is widely used to perform GSA \cite{sobol_sensitivity_1990,sobol_sensitivity_1993,saltelli_relative_2002}. We let $\p=(p_1,p_2,\dots,p_Q)^T$ denote the input model parameters and $DV(\p)$ represent the output DV summarizing an ABM simulation at the input parameter $\p$. The Sobol' Method is a variance-based approach that estimates the contribution of each model parameter $p_q$ to the model's total variance, $\Var(DV(\p))$. Sobol' introduced several sensitivity indices to quantify this contribution. In particular, we consider the \new{\textit{first-order}} and \textit{total-effect indices} in this study, which are given by \cite{smith_uncertainty_2013,saltelli_why_2019,saltelli_global_2007}:
\begin{equation}
    \new{S_{q} = \dfrac{\Var_{p_q}[\E_{\p_{\sim q}}(DV(p_q))]}{\Var(DV(\p))} \ \ q=1,2,\dots,Q, }\label{eq:first_order_index}
\end{equation}
and 
\begin{equation}
    S_{T_q}= \dfrac{\E_{\p_{\sim q}}[\Var_{p_q}(DV(\p_{\sim q}))]}{\Var(DV(\p))},\ \ q=1,2,\dots,Q, \label{eq:total_effect_index}
\end{equation}
respectively. Here, $\p_{\sim q}$ represents all parameters except for $p_q$, and $\E_{\p_{\sim q}}(\cdot)$ and $\Var_{\p_{\sim q}}(\cdot)$ represent the mean and variance over $\p$ while fixing $p_q$, respectively. \new{Briefly, $S_q$ measures the direct contribution of the $p_q$ parameter to the model's total variance, while $S_{T_q}$ measures the total contribution of the $p_q$ parameter to the variance both directly and indirectly through interactions with other parameters. Parameters with large $S_{q}$ and/or $S_{T_q}$ values can be considered sensitive and those with smaller values can be considered insensitive, although there are no clear guidelines on what constitutes a sensitive or insensitive parameter. Importantly, $S_{q}$ values are bound above by one, whereas $S_{T_q}$ values can exceed one because this index quantifies all interactions between 2 or more parameters.}

 The dimensions of $S_{q}$ and $S_{T_q}$ will match the dimension of $DV(\p)$; suppose all are of length $m$. Our final sensitivity indices are then \emph{averaged first order} and \emph{averaged total effect indices}, which we calculate by averaging over all values in $S_{q}$ and $S_{T_q}$:
\begin{equation}
    \bar{S}_{q} = \dfrac{1}{m} \sum_{i=1}^m  (S_{q})_i, \  \text{ and } \ \ \bar{S}_{T_q} = \dfrac{1}{m} \sum_{i=1}^m  (S_{T_q})_i,\ \ \ q=1,2,\dots, Q \label{eq:mean_total_effect_index}
\end{equation}

The values for $\bar{S}_{q}$ and $\bar{S}_{T_q}$ are typically estimated using quasi-Monte Carlo methods \cite{lima_bayesian_2021}. Saltelli \cite{homma_importance_1996,saltelli_making_2002,saltelli_variance_2010} proposed an efficient sampling scheme to compute the total effect indices for all parameter values. This sampling requires $N(Q+2)$ parameter samples; it is common practice to choose $N$ to be a power of 2. 

\noindent \uline{\textbf{The Saltelli Dataset:}} As a comparison between the SSRCA and Sobol' Methods, we compute a third dataset generated using Saltelli parameter sampling \cite{saltelli_sensitivity_2002}. Similar to the large 10-parameter dataset, we generate this dataset by varying the same $Q=10$ parameters and choosing $N=2^9=512$ to ensure both datasets are of comparable size: the large 10-parameter dataset contains 5,445 samples while the Saltelli Dataset contains 6,144 samples. We implemented Saltelli sampling and the Sobol' Index computations using the \textbf{SALib} package (version 1.5.1).

\section{Results}\label{sec:results}

\subsection{\new{Quantifying the baseline simulation's proliferative ring and necrotic core}}\label{sec:baseline_results}

Our baseline simulation refers to the model simulation computed using the baseline parameter values from Table \ref{table: parameters perturbed}. This simulation begins with all cells in \new{the three living states} at time 0 (Figure \ref{fig:model_overview}(c)). An outer ring of \new{proliferating} cells forms by day 2, and a necrotic core has developed by day 4. The ring and core are clearly established by day 10, and the spheroid has grown in size over the 10-day simulation.

To summarize the baseline simulation, we compute its Cell counts and Final cell density DVs. According to the Cell counts DV, there is a steady increase in the number of cells in the G1 state over the 10 day simulation, beginning with 850 and finishing with 3,000 cells, while the number of proliferating cells remains relatively constant around 10 cells in the early S state and  250 cells in the S/G2/M state (Figure \ref{fig:model_overview}(d)). The dead cell population initially grows slowly before increasing after \new{$t=4$ days}; the simulation finishes with 1,000 dead cells. The Final cell density DV appears radially symmetric with an inner ring of 137 $\mu m$ and an outer ring of about 322 $\mu m$ (Figure \ref{fig:model_overview}(e)).

\subsection{\new{SSRCA identifies four model phenotypes within the simple 2-parameter dataset}}\label{sec:dataset1_results}

We explore the two-dimensional Klowss Model's behavior when varying two parameters, $\eta_1$ and $c_a$, to generate \new{the simple 2-parameter dataset}. We summarized all \new{1,210} simulations in this dataset by computing the Cell counts and Final cell density DVs. We then reduced the dimensionality of all DVs using PCA and determined that 3-dimensional DRDVs are suitable representatives for the DVs  (Supplementary Figure \ref{fig:scree_plots}(a)-(b)) for both DV types. 

\begin{figure}
    \centering
    \includegraphics[width=0.95\linewidth]{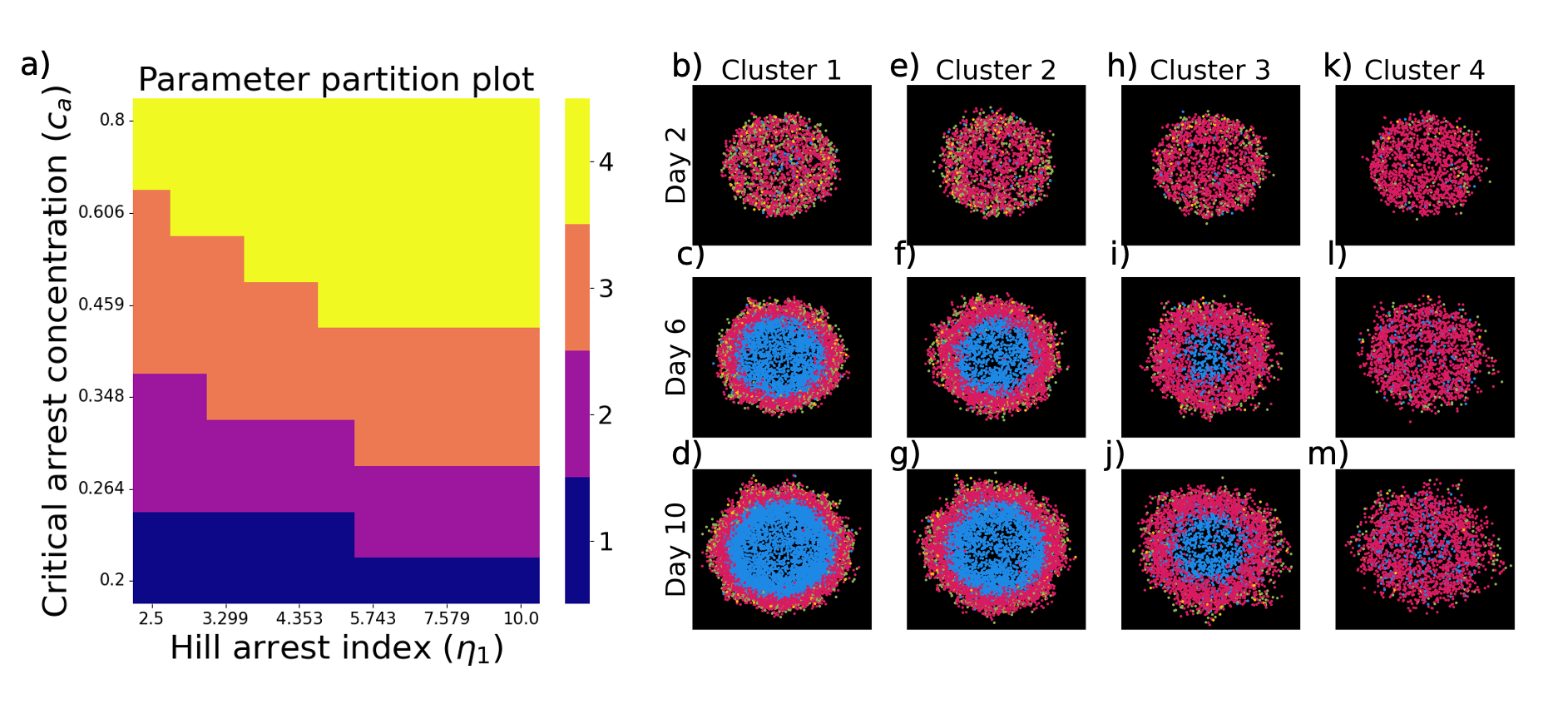}
    \caption{\new{The simple 2-parameter dataset} Analysis using the Cell counts DV. (a) The ($\eta_1, c_a$) parameter partition plot for \new{the simple 2-parameter dataset} when using the Cell counts DV. Snapshots at times $t= 2, 6, \text{ and } 10$ days from the representative model simulations from (b-d) Cluster 1, (e-g) Cluster 2, (h-j) Cluster 3, and (k-m) Cluster 4.
}
    \label{fig:DS1_counts}
\end{figure}

We trained the $k$-means clustering algorithm to the training datasets using both types of DRDVs separately. \new{Elbow plots suggest there may be either $k=4$ or $k=5$ clusters in the data (Supplementary Figure \ref{fig:elbow_plots}(a)-(b)), we found that either choice leads to similar results. We proceed with the simpler $k=4$ scenario moving forward.} When labeling each $(\eta_1,c_a)$ combination based on its most common clustering assignment, we find a clear partitioning of the $(\eta_1,c_a)$ parameter space into four distinct regions (Figure \ref{fig:DS1_counts}(a) and Supplementary Figure \ref{fig:supp_DS1_density}(a)). The parameter partition  plots that result from both types of DRDVs led to high OOS \new{consistency} values, indicating that that these parameter designations are robust (Table \ref{tab:OOS_scores}).

\begin{table}
\centering
\begin{tabular}{|l|c|c|}
\hline
&  \textbf{Cell counts DRDVs} & \textbf{Final cell density DRDVs} \\
\hline
\new{The simple 2-parameter dataset} & 99.2\% & 98.0\% \\
\new{The large 10-parameter dataset} & 99.6\% & 98.1\%  \\
\hline
\end{tabular}
\caption{Out of sample (OOS) \new{consistency} scores for both datasets when using either the Cell counts or Final cell density DRDVs. \label{tab:OOS_scores}}

\end{table}

\new{Investigation of the parameter partition plot for the Cell counts DRDVs reveals that each cluster contains only a subset of the  $c_a$ values (Figure \ref{fig:DS1_counts}(a)). In contrast, all clusters include the full range of $\eta_1$ values. These results suggest that $c_a$ impacts the model's behavior more than $\eta_1$.} Cluster 1 results for $c_a \le 0.232$, Cluster 2 results for $0.232 \le c_a \le 0.348$, Cluster 3 results for $0.306 \le c_a \le 0.606$, and Cluster 4 results for $c_a \ge 0.459$. We visualized representative model simulations from each cluster (Figure \ref{fig:DS1_counts}(b)-(m)); the simulation from Cluster 1 results in the earliest-forming and largest final necrotic core, whereas the simulation from Cluster 4 has the smallest and last-forming necrotic core. Visual inspection of the Cell counts DVs of these representative simulations reveal that as we increase the cluster assignment from 1 to 4, the number of \new{G1} cells decreases and the number of dead cells decreases (Supplementary Figure \ref{fig:DS1_representative_DVS}(a)-(d)). \new{Similarly, the Final cell density DV leads to similar representative model simulations (Supplementary Figure \ref{fig:supp_DS1_density}(b)-(m)), and the representative DVs show that the ring of living cells decreases in radius as the clustering assignment increases (Supplementary Figure \ref{fig:DS1_representative_DVS}(e)-(h)).}

We did not perform statistical analysis between the clusters from this dataset due to the ease in visually interpreting each parameter's impact in Figure \ref{fig:DS1_counts}(a) and Supplementary Figure \ref{fig:supp_DS1_density}(a). 

\subsection{\new{SSRCA identifies sensitive cell cycle- and death-related parameters from the  large 10-parameter dataset}}\label{sec:dataset2_results}

We further explore the two-dimensional Klowss Model's behavior by sampling all 10 ``Varying'' parameters from Table \ref{table: parameters perturbed} to generate \new{the large 10-parameter dataset}. We summarized all 54,450 simulations in this dataset by computing the Cell count and Final cell density DVs and found that 3-dimensional DRDVs are suitable representatives for the DVs (Supplementary Figure \ref{fig:scree_plots}(c)-(d)), and that $k=4$ reliably clusters the data (Supplementary Figure \ref{fig:elbow_plots}(c)-(d)). We trained the $k$-means clustering algorithm to the training datasets for both types of DRDVs, and the parameter labels that resulted from both types of DRDVs led to high OOS \new{consistency} values (Table \ref{tab:OOS_scores}).

\begin{sidewaysfigure}
    \centering
    \includegraphics[width=0.99\linewidth]{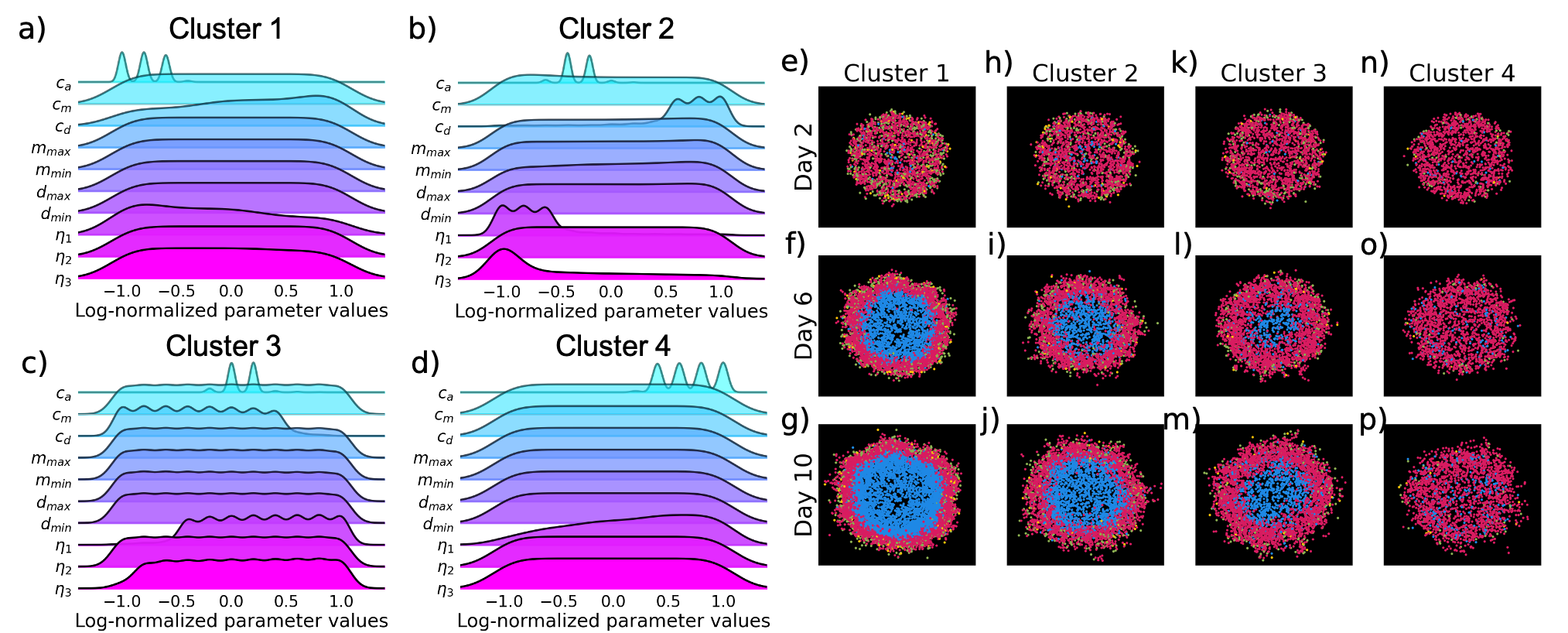}
    \caption{The \new{large 10-parameter dataset} Analysis using the Cell counts DV. (a-d) Ridgeline plots for Clusters 1-4. Each parameter $p$ is log-normalized as $\log_2(p/p_{base})$, where $p_{base}$ denotes the baseline model value from Table \ref{table: parameters perturbed}. Snapshots at times $t = 2, 6, \text{ and } 10$ days from the representative model simulations from (e-g) Cluster 1, (h-j) Cluster 2, (k-m) Cluster 3, and (n-p) Cluster 4.}
    \label{fig:DS2_counts}
\end{sidewaysfigure}

Ridgeline plots visualize the distribution of parameter samples for all four clusters. The ridgeline plots when using Cell counts (Figure \ref{fig:DS2_counts}(a)-(d)) or Final cell density DRDVs (Supplementary Figure \ref{fig:DS2_density}(a)-(d)) are similar, so we present the Cell counts DRDV results here \new{for simplicity}. \new{The distributions for the $c_a$, $c_d$, $\eta_1$, and $\eta_3$ parameters each have distinct domains between the four clusters, which indicates they are sensitive parameters. The remaining six parameters have similar domains for all four clusters and are likely insensitive. For the four sensitive parameters, we notice that:} 
\begin{itemize}
\item Cluster 1 emerges from low $c_a$ values; 
\item Cluster 2 emerges from \new{low-to-moderate} $c_a$ values, large $c_d$ values, low $\eta_1$ values, and low $\eta_3$ values; 
\item Cluster 3 emerges from \new{moderate-to-large}  $c_a$ values, low to moderate $c_d$ values, and moderate to large $\eta_1$ values; and 
\item Cluster 4 emerges from large $c_a$ values, and large $\eta_1$ values.
\end{itemize}

\new{We visualize representative simulations from each cluster (Figure \ref{fig:DS2_counts}(e)-(p)). The Cluster 1 simulation has the largest necrotic core and thinnest proliferating ring. The size of the core shrinks and the thickness of the ring increases for the simulations from Clusters 2 and 3. The Cluster 4 simulation does not include a necrotic core and consists primarily of living cells. The DVs for for each representative simulation quantify these differences (Supplementary Figure \ref{fig:DS2_representative_DVS}(a)-(d)).} \new{We find similar representative simulations when using the Final cell density DV (Supplementary Figure \ref{fig:DS2_density}(e)-(p)), and their respective DVs show that the simulations' radii decrease as the clustering assignment increases  (Supplementary Figure \ref{fig:DS2_representative_DVS}(e)-(h)).}
 
We confirmed the sensitive model parameters \new{for this dataset} by performing discrete two-sample Kolmogorov-Smirnov tests between all cluster pairs for all parameters when using the Cell counts DRDVs (Table \ref{tab:DS2_qS_cell_counts}). The tests find that the  $c_a,\ c_d,\ \eta_1, \text{ and } \eta_3$ parameters \new{achieve distributions with low $p$-values between the majority of  cluster pairs. This suggests that the distributions of these parameters are different between clusters, which we interpret as parameter sensitivity. All other parameters achieve higher $p$-values between cluster pairs, indicative of similar distributions between clusters and parameter insensitivity.} Performing the same sensitivity analysis using Final cell density DRDVs identifies these same four parameters are sensitive, as well as the $m_{min}$ parameter (Supplementary Table \ref{tab:DS2_qS_final_density}).

\begin{table}[h]
\centering
\begin{tabular}{c|cccccc|c}
& \multicolumn{6}{c|}{Cluster pairs} & \\ 
 & 1 \& 2 & 1 \& 3 & 1 \& 4 & 2 \& 3 & 2 \& 4 & 3 \& 4 & Classification \\
\hline
$c_a$ & \uline{$<10^{-3}$} & \uline{$<10^{-3}$} & \uline{$<10^{-3}$} & \uline{$<10^{-3}$} & \uline{$<10^{-3}$} & \uline{$<10^{-3}$} & Sensitive \\
$c_d$ & \uline{$<10^{-3}$} & \uline{$<10^{-3}$} & 0.059 & \uline{$<10^{-3}$} & \uline{$<10^{-3}$} & \uline{$<10^{-3}$} & Sensitive \\
$c_m$ & 0.93 & 0.9 & 0.998 & 0.241 & 0.945 & 0.874 & Insensitive \\
$d_{max}$ & 0.34 & 0.916 & 1.0 & 0.963 & 0.303 & 0.893 & Insensitive \\
$d_{min}$ & 0.786 & 0.903 & 1.0 & 0.919 & 0.804 & 0.891 & Insensitive \\
$m_{max}$ & 0.659 & 0.892 & 0.999 & 0.956 & 0.653 & 0.898 & Insensitive \\
$m_{min}$ & 0.623 & 0.877 & 1.0 & 0.944 & 0.607 & 0.894 & Insensitive \\
$\eta_1$ & \uline{$<10^{-3}$} & \uline{$<10^{-3}$} & \uline{$<10^{-3}$} & \uline{$<10^{-3}$} & \uline{$<10^{-3}$} & \uline{$<10^{-3}$} & Sensitive \\
$\eta_2$ & 0.929 & 0.87 & 1.0 & 0.843 & 0.93 & 0.891 & Insensitive \\
$\eta_3$ & \uline{ $ 0.004 $ } & \uline{ $ 0.008 $ } & 0.826 & \uline{$<10^{-3}$} & \uline{$<10^{-3}$} & \uline{$<10^{-3}$} & Sensitive \\
\end{tabular}
\label{tab:DS2_qS_cell_counts}
\caption{Determining sensitive  parameters from \new{the large 10-parameter dataset} when using the Cell counts DRDV. \new{A small $p$-value between two distributions indicates they are significantly different from each other.} We designate $p$-values below 0.05 as being significant. If a parameter is significantly different for the majority of cluster pairs, then we classify the parameter as sensitive.}
\end{table}

\subsection{\new{SSRCA provides stable sensitivity analysis results while Sobol' does not}}

\begin{figure}[h!]
    \centering
    \includegraphics[width=0.9\linewidth]{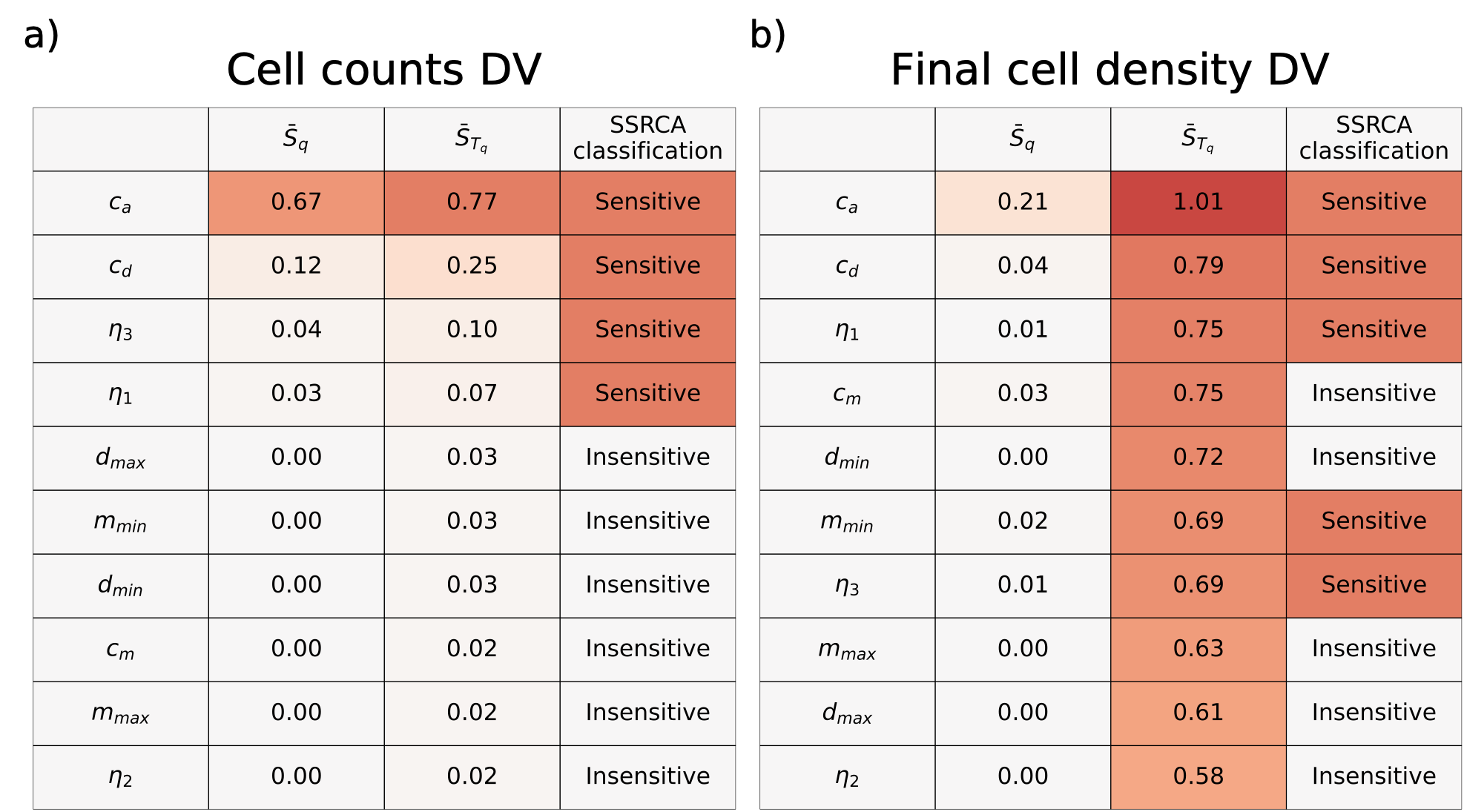}
    \caption{Comparing Sobol' and SSRCA sensitivity analysis. We performed sensitivity analysis for the 10 model parameter using the Sobol' indices methodology and the SSRCA Methodology \new{when using the (a) Cell counts DV and (b) the Final cell density DV}.}
    \label{fig:sobol_comparison}
\end{figure}

We compared the sensitivity analysis results from applying the SSRCA and Sobol' Methods \new{to Cell count DVs and Final cell density DVs}. For the Sobol' Method, we computed the averaged \new{first-order} and total-effect indices, given by $\bar{S}_{q}$ and $\bar{S}_{T_q}$, respectively, from Equation \eqref{eq:mean_total_effect_index}, when varying all 10 parameters to generate the Saltelli Dataset. \new{The SSRCA and Sobol' Methods produce similar sensitivity analysis results when using the Cell counts DVs (Figure \ref{fig:sobol_comparison}(a)):  SSRCA identifies four sensitive parameters:$c_a, c_d, \eta_3, \text{ and }\eta_1$. Each of these parameters yield the four highest $\bar{S}_{q}$ and $\bar{S}_{T_q}$ values, where $c_a$ has the highest values ($\bar{S}_{q}=0.67$, $\bar{S}_{T_q}=0.77$) followed by $c_d$ ($\bar{S}_{q}=0.12$, $\bar{S}_{T_q}=0.25$). The sensitivity index values are notably smaller for $\eta_3$ ($\bar{S}_{q}=0.04$, $\bar{S}_{T_q}=0.10$) and $\eta_1$ ($\bar{S}_{q}=0.03$, $\bar{S}_{T_q}=0.07$).}


\new{There is less agreement between the two methods when using the Final cell density DV (Figure \ref{fig:sobol_comparison}(b)). The Sobol' Method leads to all ten parameters achieving $\bar{S}_{T_q}$ values of 0.58 or greater, which could suggest that all parameters are sensitive. However, inspection of the $\bar{S}_{q}$ values reveals that only the $c_a$ parameter achieves a sensitivity index value above 0.05, while all others are at 0.04 or below. The large discrepancy between the $\bar{S}_{q}$ and $\bar{S}_{T_q}$ index values for the nine other parameters suggests that the model variance is highly dependent on interactions between parameters. Using the SSRCA Methodology with the Final cell density DVs, on the other hand, designates five total sensitive parameters: $c_a, c_d, \eta_3, \eta_1$, and $m_\text{min}$. Importantly, these are the same parameters that SSRCA designated for the Cell counts DV with the addition of the $m_\text{min}$ parameter. This fifth parameter may be a result of the Final cell density DV capturing a spatial aspect of the model that is missing from the Cell counts DV.}

\section{Discussion and conclusions}\label{sec:discussion}

We introduce the \textbf{S}imulate, \textbf{S}ummarize, \textbf{R}educe dimensionality, \textbf{C}luster, and statistically \textbf{A}nalyze (SSRCA) method as a new approach to perform sensitivity analysis (SA) for ABMs. The pipeline consists of summarizing ABM simulations with summary descriptor vectors (DVs) and then applying machine learning tools (dimensionality reduction and clustering) to these DVs to identify common model output patterns. The method concludes with statistical analysis between the clusters to interpret and identify which model parameter values most strongly influence these patterns. 
An important advantage of SSRCA over other common sensitivity analysis (SA) methods is its ability to simultaneously 1) determine sensitive parameters, 2) identify common model patterns, and 3) perform feature mapping for complex ABMs, even for a large number of parameters.

Many methodologies exist to perform SA; two of the most widely-used approaches include the Morris and Sobol' Methods \cite{morris_factorial_1991,sobol_sensitivity_1990,sobol_sensitivity_1993,ten_broeke_which_2016}. We compare the benefits and drawbacks of SSRCA with these approaches in Table \ref{tab:discussion_comparing_SAs}. All three methods can determine sensitive parameters, however, the Sobol' Method fails to identify common model patterns because it computes sensitivity indices by aggregating results over the entire parameter space. The Morris method can identify common model patterns by varying one parameter at a time and tracking how the outputs change. However, the Morris method cannot perform extensive feature mapping for these patterns because it ignores parameter interactions. A strength of the Sobol' Method is that it accounts for these interaction effects when computing its sensitivity indices. We find that the SSRCA Method can both identify model patterns and perform exhaustive feature mapping. Furthermore, we can vary multiple parameters at once to incorporate interaction effects into this analysis. A key benefit of the Morris method is its computational efficiency; previous studies estimate that the cost of the Morris method scales as $\sim 10$ times the number of parameters, whereas the cost of the Sobol' Method scales as $\sim 10^3$ times the number of parameters \cite{bergman_efficient_2025, morris_factorial_1991,marino_methodology_2008}. As a first pass, we implemented SSRCA on a dense set of parameters where we varied 10 parameters, 2 at a time. The SSRCA Method can be used with any parameter sampling, however, and in future work we plan to assess how the number of required simulations scales with the number of parameters. We will further examine its use with more efficient parameter sampling techniques, including Latin Hypercube sampling \cite{marino_methodology_2008}.

\begin{table}[]
    \centering
    \begin{tabular}{|l|c|c|c|}
    \hline
          & Morris Method & Sobol' Method & SSRCA Method \\
    \hline

        Parameter sensitivity & yes & yes &yes \\
        \makecell[l]{\new{Identify common model} \\ \new{patterns}} & yes & no & yes \\
        Exhaustive feature mapping & no & no & yes \\

        Interaction effects & no & yes & yes \\
        Computationally efficient & yes & no & \makecell{further investigation \\ required} \\

     \hline
    \end{tabular}
    
    \caption{Comparing various capabilities of the Morris, Sobol', and SSRCA Methods in performing sensitivity analysis. }
    \label{tab:discussion_comparing_SAs}
\end{table}

Another strength of SSRCA is its flexibility with the choice of methodologies used within the pipeline. We found similar results in this work whether we used the Cell counts DV or Final Cell Density DV. \new{This is in contrast to the Sobol' Method, whose sensitivity analysis results were significantly different when using the Cell counts or Final cell density DVs, even though the underlying model simulations were unchanged.} This robustness to DV choice is crucial, as ABMs are broadly used for biological problems, and the appropriate DV choices depends on the application at hand. For example, many studies rely on order parameters to characterize ABM behavior; alternatively, topological data analysis (TDA) is a recent field of research that has found success in summarizing many types of ABMs with applications in complex systems \cite{bhaskar_analyzing_2019, bhaskar_topological_2023, gharooni-fard_computational_2024}. SSRCA can be used with any of these DVs, as it only requires they are uniform length. Additionally, we used principal components analysis and $k$-means clustering for dimensionality reduction and clustering, respectively, in this work, but any suitable algorithm can be used for these processes.

We applied the SSRCA Methodology to a 2-dimensional ABM of tumor spheroid growth and identified the $c_a$, $c_d$, $\eta_1$, $\eta_3$ parameters as the most sensitive ones. \new{These parameters make biological sense as sensitive parameters because they control cell death and entry into the cell cycle. These results  may inform the future use of tumor spheroid models to interpret experimental data by focusing on these processes.} For example, the 3-dimensional Klowss Model was calibrated to experimental data from human primary melanoma cells \cite{4D_Tumor_Model}; modeling similar data from other cell lines will likely require different parameter values. Our work proposes that thoroughly exploring all unknown parameters is not necessary; instead, one may more efficiently vary the four identified sensitive parameters to fit the model to new data. Our proposed methodology may be of use for other modeling frameworks. For example, the authors of \cite{jin_mathematical_2021} could not perform parameter estimation due to their differential equation model containing 19 unknown parameters. The SSRCA Methodology could be applied to this model to learn an informative subset of its parameters for parameter estimation.

\new{There are limitations associated with the SSRCA algorithm and our implementation of it in this work.} While we can use the SSRCA Methodology to identify sensitive parameters, it is currently not able to rank parameters based on their sensitivity. We plan to further analyze the model and/or develop metrics that can be used in combination with SSRCA to enable such parameter ranking. We also demonstrated the utility of the SSRCA Method when applied to informative DVs, however, it has been reported previously that some DVs are more informative than others depending on the context \cite{bhaskar_analyzing_2019,angio_tda,nardini_statistical_2023,topaz_topological_2015}. It would be interesting to explore if the SSRCA algorithm can be used to compare the information content between various DVS, or even to propose an optimal DV choice. \new{This may be especially relevant when applying the SSRCA Method to more complex ABMs in the future. We applied the SSRCA Methodology to a simple radially symmetric and two-dimensional tumor spheroid ABM in this work; these simplifications yield simulations with similar output DVs when parameter values are kept constant. Applying our proposed methods to more complicated ABMs (e.g., radially asymmetric and/or 3-dimensional models) in the future will lead to more complex and noisy DVs, and it is not clear how SSRCA will perform in the face of these challenges or if the most sensitive parameters will be changed. We plan to perform consistency analyses to determine how many model simulations from each parameter combination are necessary to mitigate these increased uncertainly levels \cite{gerstman_uncertainty_2020}. It will also be of interest to apply the SSRCA Methodology to ABMs from other biological contexts, including cell biology, epidemiology, and ecology.}

\textbf{Data and code availability:} All code to simulate this study's data and perform data analysis are publicly available at 
\href{https://github.com/e-rohr/FUCCI_ABM2D}{https://github.com/e-rohr/FUCCI\_ABM2D}.

\textbf{Acknowledgements:} The authors acknowledge use of the ELSA high performance computing cluster at The College of New Jersey for conducting the research reported in this paper. This cluster is funded in part by the National Science Foundation under grant numbers OAC-1826915 and OAC-2320244. The authors thank Matthew Simpson for useful discussion on this study.

\bibliographystyle{unsrt}
\bibliography{tumor_spheroids.bib}  

\appendix

\renewcommand{\thesection}{S\arabic{section}}
\renewcommand{\thetable}{S\arabic{table}}
\renewcommand{\thefigure}{S\arabic{figure}}
\renewcommand{\theequation}{S\arabic{equation}}

\section{Model overview}\label{app:model_overview}

We simulate an agent-based model of tumor spheroid growth developed in \cite{4D_Tumor_Model}. This model is hybrid, meaning we track both discrete agent locations and a continuous nutrient concentration (e.g., oxygen) within the spheroid. Tumor cells are represented as discrete agents centered at $\textbf{x}_n(t) = (x_n(t),y_n(t))$
for $n = 1, \ldots , N(t)$, where $N(t)$ is the total number of agents at time $t$. The spatio-temporal distribution of the nutrient concentration is non-dimensionalized and represented by $c(\textbf{x}, t)\in [0,1]$. In addition to agent locations, we also model how cells progress through the cell cycle. The agents are colored according to FUCCI: agents in the G1 phase are colored red, agents in the early S (eS) phase are colored yellow, and agents in the G2/S/M phase are colored green. Following \cite{4D_Tumor_Model}, we assume the non-dimensional, local nutrient concentration, $c(\textbf{x},t)$, regulates cell motility, proliferation, and death \cite{greenspan_models_1972,Haass,ward_mathematical_1997}. The model is simulated in a cubic domain, $\Omega$, with side length $1000 ~\mu\mathrm{m}$, which we chose to mitigate computational costs while ensuring that agents do not reach the domain boundary during simulation.

We now describe how tumor cells in the model spheroid undergo mitosis, cell motility, and cell death, and then detail how the nutrient spreads and is consumed. All model parameters are described in Table \ref{table: parameters perturbed}. The tumor cell processes are modelled as discrete events using the Gillespie Algorithm \cite{Gillespie}. We initialize the ABM in a similar manner as \cite{4D_Tumor_Model}, where initial cell colors are randomized to produce proportions of red, yellow, and green cells that match experimental observations.

\subsection{Agent Dynamics}
We now discuss how agents progress through the cell-cycle. Each agent progresses at a rate determined by its cell cycle status. The red-to-yellow transition rate, $R_r(c)$, depends on the local nutrient concentration $c(\textbf{x},t),$ whereas the yellow-to-green transition rate, $R_y$, and the green-to-red transition rate, $R_g$, are independent of $c(\textbf{x},t)$.  The the agent cell cycle progression rates are given by
\begin{align}
    R_r(c) &= R_r\frac{c^{\eta_1}}{c_a^{\eta_1} + c^{\eta_1}},\label{eq:R->Y}\\
    R_y(c) &= R_y,\\
    R_g(c) &= R_g,
\end{align}
 (See Table \ref{table: parameters perturbed} for a description of all parameters). The cell-cycle transition from green to red includes mitosis, where a green agent produces two red agents. During this process, a polar angle $\theta$ is sampled  to obtain a random direction where the daughter agents are placed. The first daughter agent is placed a distance $\mu/2$ in the randomly chosen direction, and the second daughter agent is placed $\mu/2$ units in the opposite direction, so that the two daughter cells are a distance $\mu$ apart The step length $\mu$ is chosen to be a typical cell diameter.

We now describe the model rules on agent migration and death. The agent migration and death rates, $m(c)$ and $d(c)$, respectively, are determined by the local nutrient concentration. Motility increases with $c$ since oxygen supports energy production; whereas, the death rate decreases with $c$ as cells are less likely to asphyxiate in high oxygen levels. These rates are given by
\begin{align}
    m(c) &= (m_{max} - m_{min})\frac{c^{\eta_2}}{c_m^{\eta_2} + c^{\eta_2}} + m_{min},\label{eq:move}\\
    d(c) &= (d_{max} - d_{min})\left(1-\frac{c^{\eta_3}}{c_d^{\eta_3} + c^{\eta_3}}\right) + d_{min}\label{eq:death},
\end{align}
where all parameters are described in Table $\ref{table: parameters perturbed}$. When migrating, an agent is displaced $\mu$ units in the randomly chosen direction. \new{Following \cite{4D_Tumor_Model}, cellular interactions are ignored during cell migration, cell cycle progression and death; \textit{i.e.}, these process are unaffected by cell mechanics, crowding, and/or cell-cell interactions.} When an agent dies, it is removed from the simulation by setting its status to dead, and its final location is recorded.

\subsection{Nutrient dynamics}
\label{Nutrient dynamics}
The spatio-temporal distribution of the dimensionalized nutrient concentration, $C(\textbf{x},t)$, is assumed to be governed by the reaction-diffusion equation
\begin{equation}
\label{eq: nutrient reaction-diffusion}
    \frac{\partial C}{\partial t} = D\nabla^2C - \kappa C v, \text{ in }\Omega,
\end{equation}
with diffusivity $D > 0 ~ [\mu\mathrm{m}^2 ~ \mathrm{h}^{-1}]$, consumption rate $\kappa > 0~ [\mu m^2 ~(\mathrm{h}~ \text{cells})^{-1}]$, and where $v(\textbf{x},t)\geq 0 ~[\text{cells }\mu m^{-2}]$ is the density of agents at position $\textbf{x}$ and time $t.$ On the boundary, $\partial\Omega$, a maximum far-field concentration $C(\textbf{x},t) = c_b$ is imposed. We non-dimensionalize this independent variable by setting $c(\textbf{x},t) = C(\textbf{x},t)/c_b$, so that $c = 1$ on $\partial\Omega$, and $c(\textbf{x},t)\in [0,1]$. We approximate the reaction-diffusion Equation \eqref{eq: nutrient reaction-diffusion} using the quasi-steady state equation
\begin{equation}
    \label{eq: nutrient quasi-steady}
    0 = \nabla^2c - \alpha c v, \text{ in }\Omega,
\end{equation}
where $\alpha = \kappa/D > 0\ [\mu \mathrm{m} ~ \mathrm{cell}^{-1}]$. During simulations, the spatio-temporal distribution of the chemical is modelled by solving Equation \eqref{eq: nutrient quasi-steady} $t^*$ units using a finite volume method on a uniform structured mesh with node spacing $h$. Following \cite{4D_Tumor_Model}, we use $t^*=1$. The cell density $v(\textbf{x},t)$ is approximated by setting $v(\textbf{x}_{i,j},t) = N_{i,j}/h^2$, where $N_{i,j}$ is the number of agents within the control volume surrounding the node located at $(x_i,y_j)$ and $h^2$ is the volume of the control volume.

\subsection{Initializing the ABM}\label{initialization}

\new{We initialize model simulations according to the same rules from \cite{4D_Tumor_Model}. We sample the $N(0)$ initial cell locations in polar coordinates. For the radius of the initial locations, we sample $U$ for each cell from the uniform distribution between 0 and 1. We then obtain each cell's initial radius by $R\sqrt{U}$, where $R$ is the maximum initial radius ($[0,245]$ $\mu$m). Each cell's initial angle is uniformly sampled between $[0,2\pi]$ radians. The phase of each cell is then determined by the value of $c(\boldsymbol{x},0)$ at its initial location. Cells with initial locations satisfying $c(\boldsymbol{x},0)>c_a$ are in a ``freely-cycling region'' where 53.92\% are in the G1 phase, 5.06\% are in the early S phase, and 41.01\% are in the S/G2/M phase.  Cells with initial locations satisfying $c(\boldsymbol{x},0)\le c_a$ are in a ``region of restricted nutrient concentration'' where 84\% of cells are in the G1 phase, 10.99\% are in the early S phase, and 89.01\% are in the S/G2/M phase. These percentages for each region were chosen to match measured cell proportions \cite{4D_Tumor_Model}. }

\section{Interpretation of ridgeline plots}\label{app:ridgeline}

\begin{figure}
    \centering
    \includegraphics[width=0.85\linewidth]{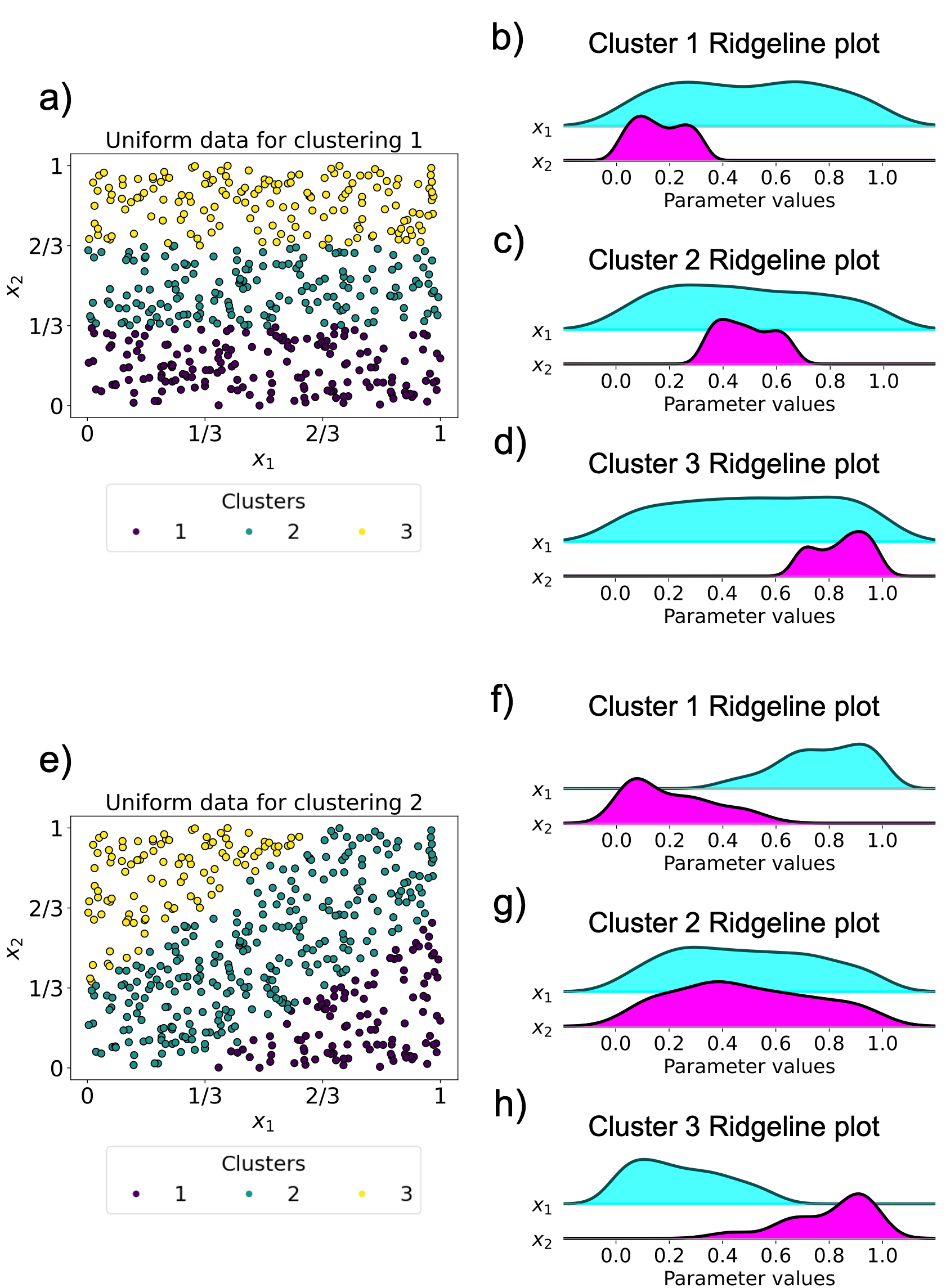}
    \caption{Ridgeline interpretation plot. (a) Uniformly sampled data $(x_1,x_2)$ where only $x_2$ is sensitive to the cluster labeling. (b)-(d) Ridgeline plots for Clusters 1-3 for this labeling.  (e) Uniformly sampled data $(x_1,x_2)$ where $x_1$ and $x_2$ interact for the data cluster labeling. (f)-(h) Ridgeline plots for Clusters 1-3 for this labeling.}
    \label{fig:supp_ridgeline_interpretation}
\end{figure}

\new{In this section, we discuss how to interpret sensitive and insensitive parameters from ridgeline plots over various data clusters. We created a simple dataset by performing 500 samples for two parameters, $x_1\text{ and }x_2$, from a uniform distribution on the domain $[0,1]$. We will illustrate how to identify (a) sensitive vs. insensitive parameters and (b) parameter interactions through two example clusterings of this data. In clustering 1, We designate each sample $(x_1,x_2)$ according to the following rule: samples with $x_2\in[0,1/3]$ belong to cluster 1, samples with $x_2\in(1/3,2/3]$ belong to cluster 2, and samples with $x_2\in(2/3,1]$ belong to cluster 3 (Supplementary Figure \ref{fig:supp_ridgeline_interpretation}(a)). Here, only $x_2$ determines a sample's clustering assignment (so it is sensitive) and $x_1$ has no impact (so it is insensitive). From this data and clustering, the ridgeline plot is constructed as follows. For each cluster, we determine all values for $x_1$ and $x_2$ within that cluster and plot these marginal distributions for each parameter (Supplementary Figure \ref{fig:supp_ridgeline_interpretation}(b-d)). The sensitivity of $x_2$ is shown through the three peaked distributions for $x_2$ in the ridgeline plots. The insensitivity of $x_1$ is shown through the three widely-distributed distributions for $x_1$ in the ridgeline plots. }

\new{We created data clustering 2 to demonstrate how ridgeline plots show parameter interactions between clusters. In this clustering, we place samples where $x_2-x_1 \le -1/3 $ into cluster 1, samples where $-1/3 \le x_2-x_1 \le 1/3 $ into cluster 2, and samples where $x_2-x_1 \le 1/3 $ into cluster 3 (Supplementary Figure \ref{fig:supp_ridgeline_interpretation}(e)). Here, the parameters interact in generating the clusterings, as cluster 1 contains samples with large values of $x_1$ and small values of $x_2$, cluster 2 contains samples with all values of $x_1$ and $x_2$, and  cluster 3 contains samples with small values of $x_1$ and large values of $x_2$. These various distributions are represented in the ridgeline plots for all clusters (Supplementary Figure \ref{fig:supp_ridgeline_interpretation}(f-h)). }

\begin{figure}
    \centering
    \includegraphics[width=0.95\linewidth]{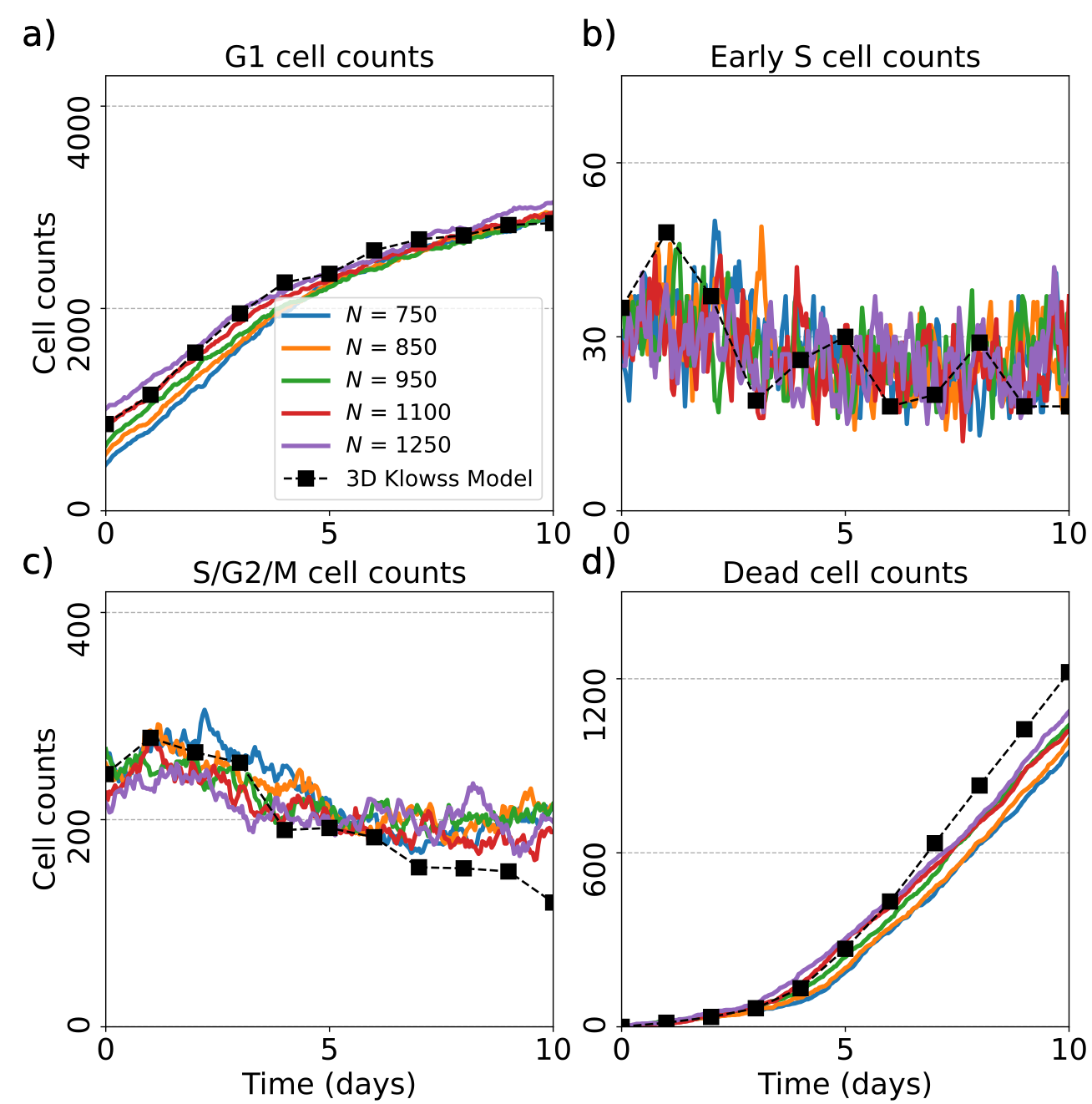}
    \caption{Determining the baseline $N(0)$ value. To determine the baseline value of $N(0)$, we visually compared simulations of the 2-dimensional Klowss model (colored lines) for various values of $N(0)$ to the $z=0$ cross section of the 3-dimensional Klowss Model from \cite{4D_Tumor_Model} (black squares) for cells in the (a) G1 state, (b) Early S state, (c) S/G2/M state, and (d) dead state .}
    \label{fig:supp_determine_N}
\end{figure}

\begin{figure}
    \centering
    \includegraphics[width=0.95\linewidth]{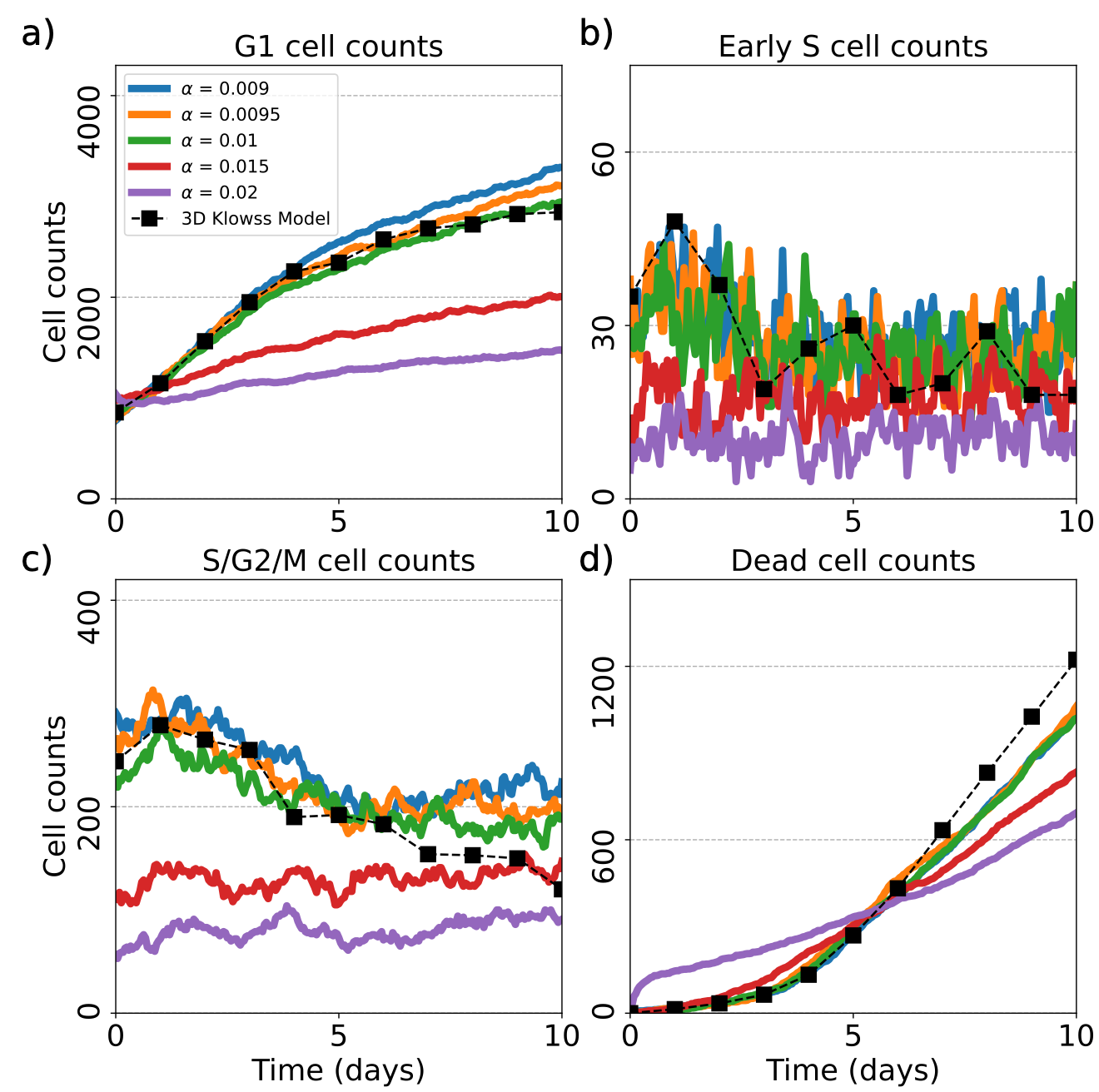}
    \caption{Determining the baseline $\alpha$ value. To determine the baseline value of $\alpha$, we visually compared simulations of the 2-dimensional Klowss model (colored lines) for various values of $\alpha$ to the $z=0$ cross section of the 3-dimensional Klowss Model from \cite{4D_Tumor_Model} (black squares) for cells in the (a) G1 state, (b) Early S state, (c) S/G2/M state, and (d) dead state .}
    \label{fig:supp_determine_alpha}
\end{figure}

\begin{figure}
    \centering
    \includegraphics[width=0.95\linewidth]{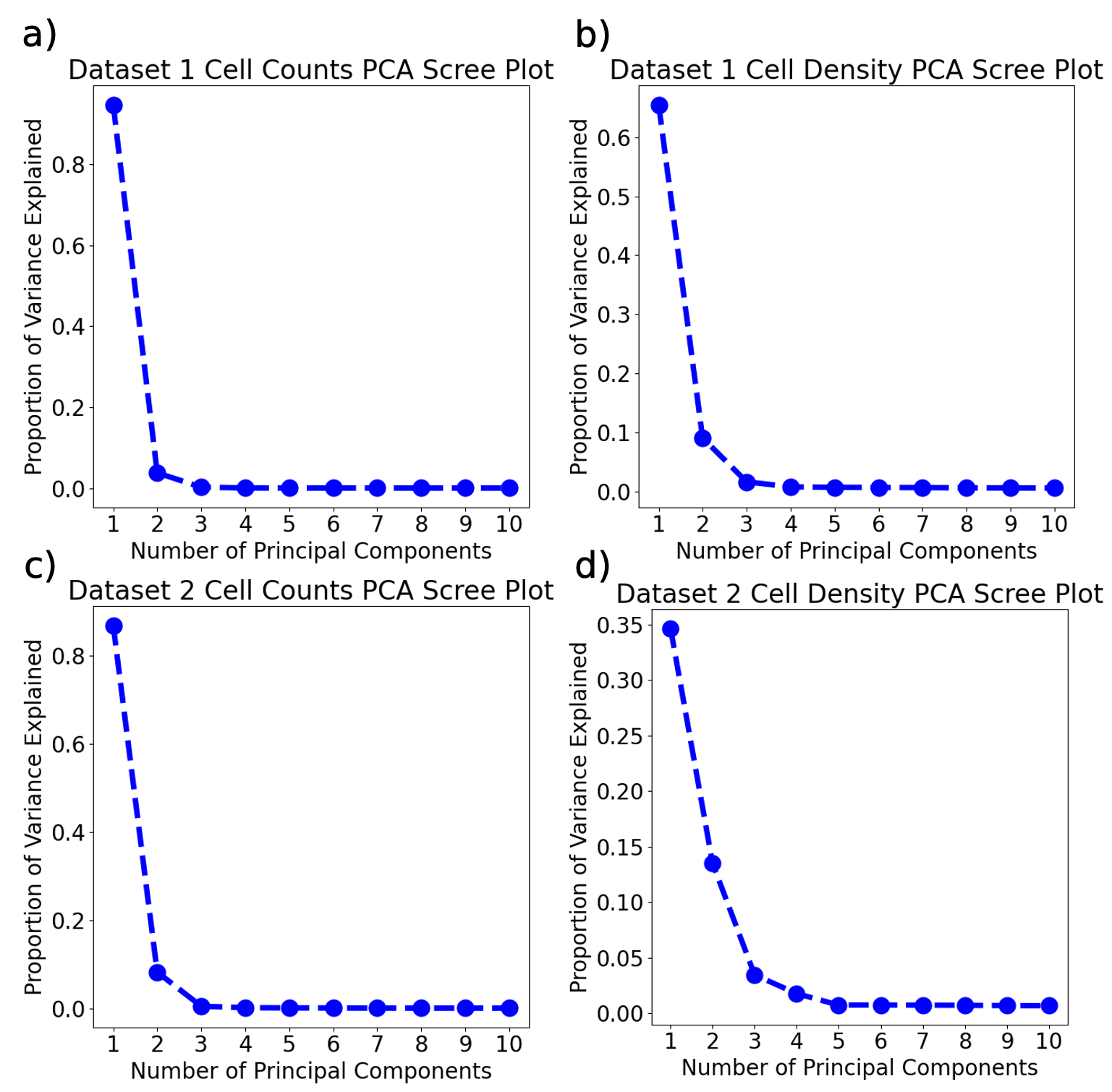}
    \caption{Scree plots. The scree plots that result when performing dimensionality reduction using (a) the small 2-parameter dataset using the Cell counts DRDV, (b) the small 2-parameter dataset using the Final cell density DRDV, (c) the large 10-parameter dataset using the Cell counts DRDV, (d) the large 10-parameter dataset using the Final cell density DRDV}
    \label{fig:scree_plots}
\end{figure}

\begin{figure}
    \centering
    \includegraphics[width=0.95\linewidth]{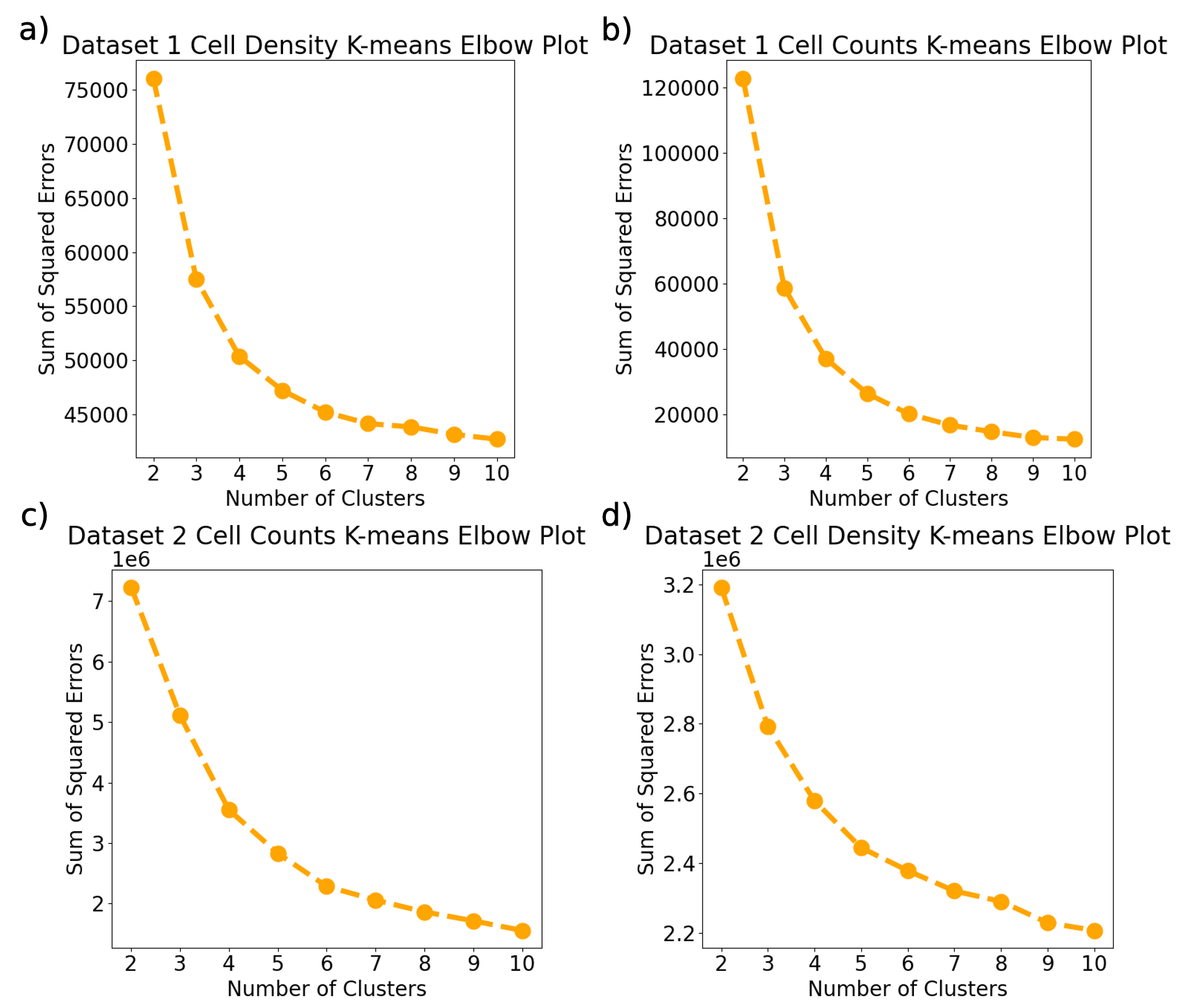}
    \caption{Elbow plots. The elbow plots that result when clustering data from (a) the small 2-parameter dataset using the Cell counts DRDV, (b) the small 2-parameter dataset using the Final cell density DRDV, (c) the large 10-parameter dataset using the Cell counts DRDV, (d) the large 10-parameter dataset using the Final cell density DRDV}
    \label{fig:elbow_plots}
\end{figure}

\begin{figure}
    \centering
    \includegraphics[width=0.95\linewidth]{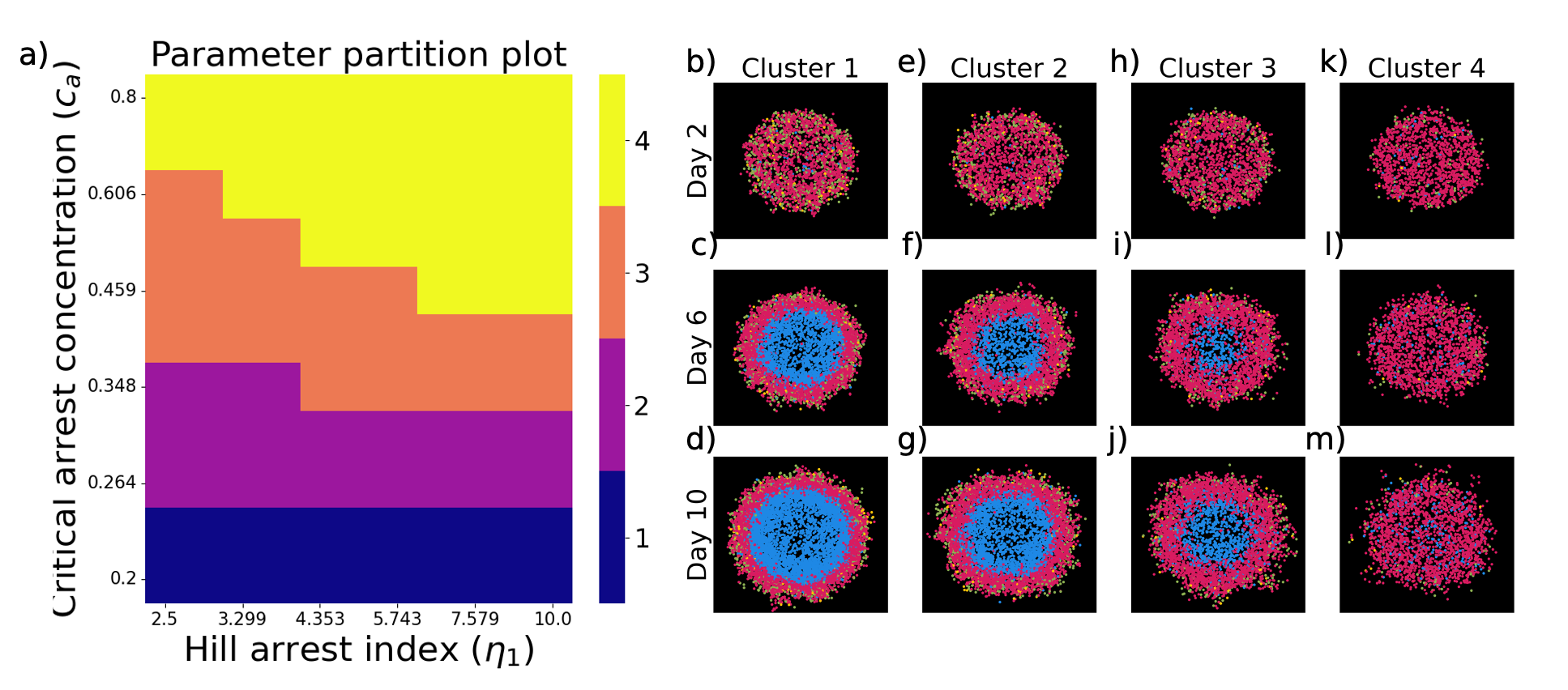}
    \caption{The small 2-parameter dataset Analysis using the Final cell density DV. (a) The ($\eta_1, c_a$)  Parameter partition plot for the small 2-parameter dataset when using the Cell counts DV. Snapshots at times $t=2, 6, \text{ and } 10$ days from the representative model simulations from (b-d) Cluster 1, (e-g) Cluster 2, (h-j) Cluster 3, and (k-m) Cluster 4. }
    \label{fig:supp_DS1_density}
\end{figure}


\begin{figure}
    \centering
    \includegraphics[width=0.95\linewidth]{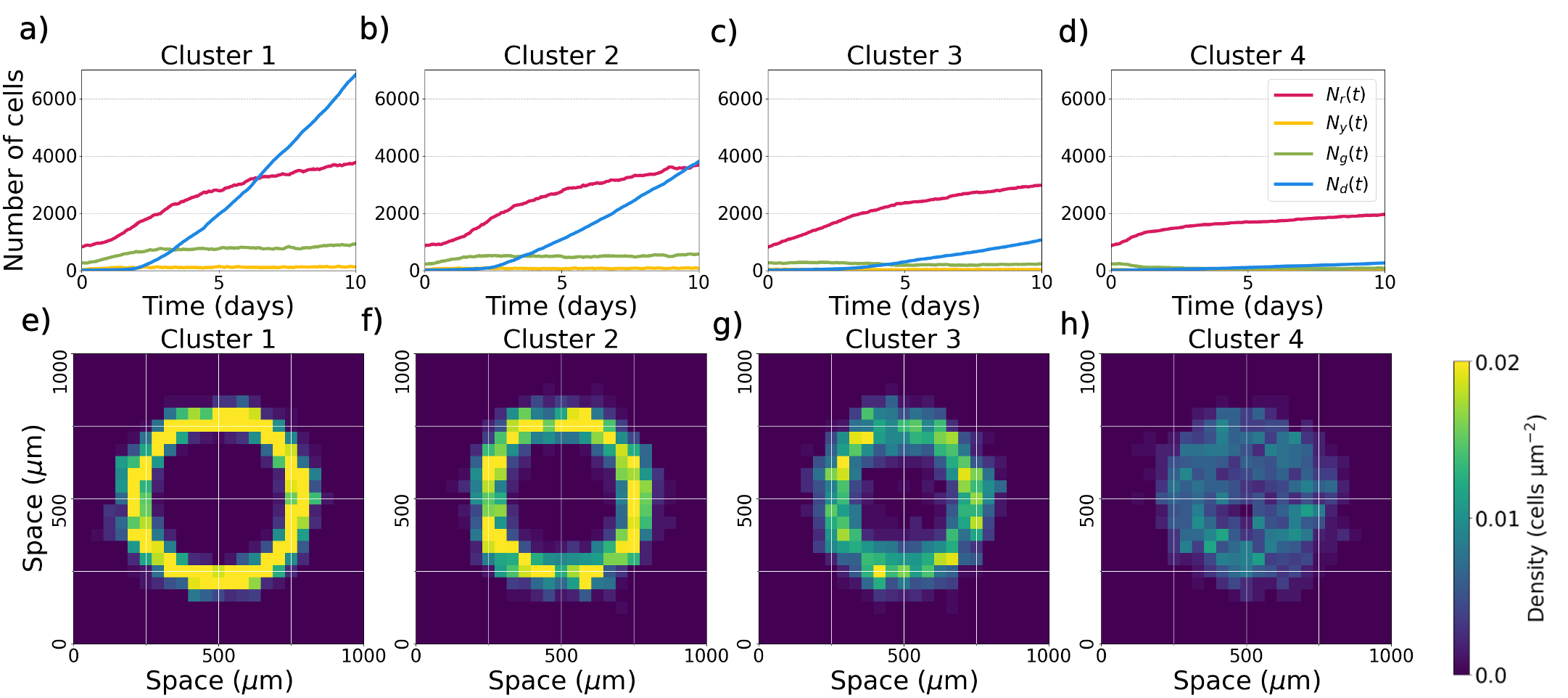}
    \caption{Represenative DVs for \new{the small 2-parameter dataset}. Representative Cell counts DVs for the small 2-parameter dataset from (a-d) clusters 1-4. Representative Final cell density DVs for the small 2-parameter dataset from (e-h) clusters 1-4.}
    \label{fig:DS1_representative_DVS}
\end{figure}

\begin{sidewaysfigure}
    \centering
    \includegraphics[width=0.99\linewidth]{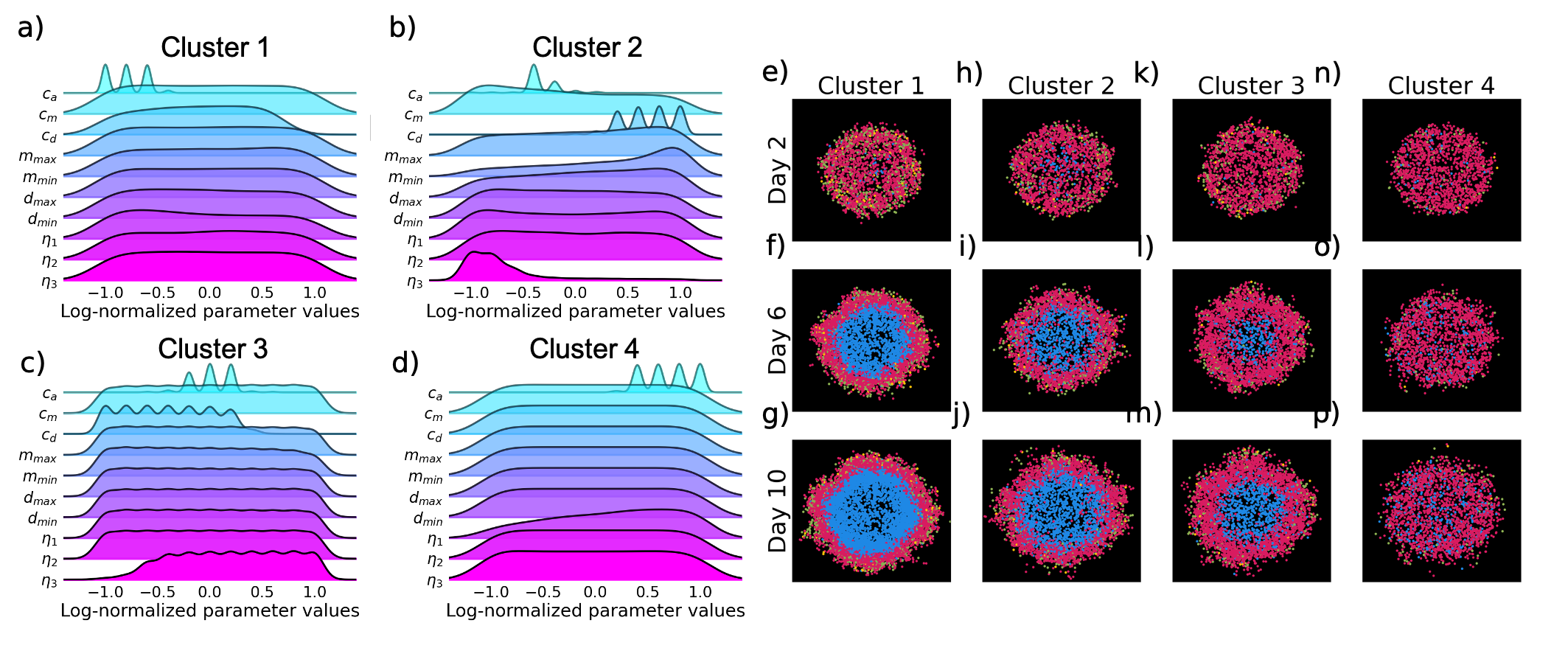}
    \caption{The large 10-parameter dataset Analysis using the Final cell density DV. (a-d) Ridgeline plots for Clusters 1-4. Each parameter $p$ is log-normalized as $\log_2(p/p_{base})$, where $p_{base}$ denotes the baseline model value from Table \ref{table: parameters perturbed}. Snapshots at times $t = 2, 6, \text{ and } 10$ days from the representative model simulations from (e-g) Cluster 1, (h-j) Cluster 2, (k-m) Cluster 3, and (n-p) Cluster 4.}
    \label{fig:DS2_density}
\end{sidewaysfigure}

\begin{figure}
    \centering
    \includegraphics[width=0.95\linewidth]{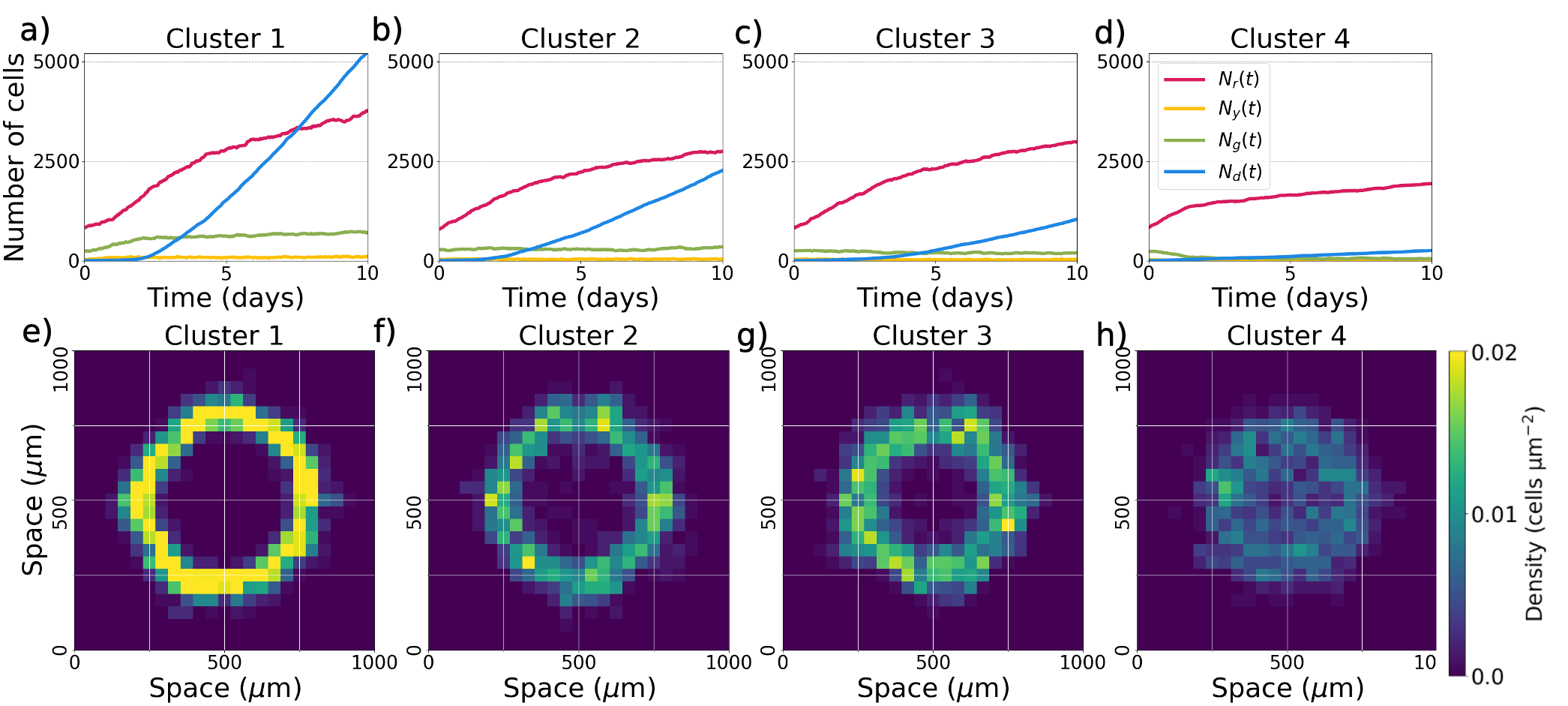}
    \caption{Represenative DVs for \new{the large 10-parameter dataset}. Representative Cell counts DVs for the large 10-parameter dataset from (a-d) clusters 1-4. Representative Final cell density DVs for the large 10-parameter dataset from (e-h) clusters 1-4.}
    \label{fig:DS2_representative_DVS}
\end{figure}


\begin{table}[h]
\centering
\label{tab:DS2_qS_final_density}
\begin{tabular}{c|cccccc|c}
& \multicolumn{6}{c|}{Cluster pairs} & \\ 
 & 1 \& 2 & 1 \& 3 & 1 \& 4 & 2 \& 3 & 2 \& 4 & 3 \& 4 & Classification \\
 \hline
$c_a$ & \uline{$<10^{-3}$} & \uline{$<10^{-3}$} & \uline{$<10^{-3}$} & \uline{$<10^{-3}$} & \uline{$<10^{-3}$} & \uline{$<10^{-3}$} & Sensitive \\
$c_d$ & \uline{$<10^{-3}$} & \uline{ $ 0.008 $ } & \uline{$<10^{-3}$} & \uline{$<10^{-3}$} & \uline{$<10^{-3}$} & \uline{$<10^{-3}$} & Sensitive \\
$c_m$ & 0.307 & 0.626 & 0.995 & \uline{$<10^{-3}$} & 0.098 & 0.712 & Insensitive \\
$d_{max}$ & \uline{ $ 0.048 $ } & 0.915 & 0.999 & 1.0 & \uline{ $ 0.027 $ } & 0.911 & Insensitive \\
$d_{min}$ & 0.767 & 0.514 & 0.883 & 0.525 & 0.961 & 0.875 & Insensitive \\
$m_{max}$ & \uline{ $ 0.023 $ } & 0.884 & 0.999 & 0.998 & \uline{ $ 0.016 $ } & 0.9 & Insensitive \\
$m_{min}$ & \uline{$<10^{-3}$} & 0.964 & 1.0 & \uline{$<10^{-3}$} & \uline{$<10^{-3}$} & 0.911 & Sensitive \\
$\eta_1$ & \uline{ $ 0.048 $ } & 0.061 & \uline{$<10^{-3}$} & 0.215 & \uline{ $ 0.01 $ } & 0.246 & Sensitive \\
$\eta_2$ & 0.934 & 0.784 & 0.998 & 0.361 & 0.975 & 0.878 & Insensitive \\
$\eta_3$ & \uline{$<10^{-3}$} & \uline{$<10^{-3}$} & 1.0 & \uline{$<10^{-3}$} & \uline{$<10^{-3}$} & \uline{$<10^{-3}$} & Sensitive \\
\end{tabular}
\caption{Determining sensitive the large 10-parameter dataset parameters from the Final cell density DRDV. We designate $p$-values below 0.05 as being significant. If a parameter receives $p$-values below 0.05 for the majority of cluster pairs, then we classify the parameter as sensitive.}
\end{table}

\end{document}